\documentclass[11pt]{article}
\usepackage{graphicx}
\usepackage{epsfig}

\usepackage{amsfonts,amssymb,amsmath}
\usepackage{latexsym}
\usepackage[english]{babel}

\newcommand{\eq}{\begin{equation}}
\newcommand{\feq}{\end{equation}}
\newcommand{\eeq}{\end{equation}}
\newcommand{\eqn}{\begin{eqnarray}}
\newcommand{\feqn}{\end{eqnarray}}
\newcommand{\arr}{\begin{eqnarray*}}
\newcommand{\farr}{\end{eqnarray*}}

\newcommand{\A}{{\cal A}}

\newcommand{\calP}{{\cal P}}

\newcommand{\calS}{{\cal S}}
\newcommand{\T}{{\cal T}}

\newcommand{\dP}{{\partial\calP}}
\newcommand{\intP}{\int_{\calP}}

\newcommand{\eg}{{\it e.g.,}\ }
\newcommand{\ie}{{\it i.e.,}\ }
\newcommand{\p}{\partial}

\newcommand{\lp}{\left(}
\newcommand{\rp}{\right)}
\newcommand{\lpp}{\left[}
\newcommand{\rpp}{\right]}
\newcommand{\ri}{R_\mathrm{i}}
\newcommand{\ro}{R_\mathrm{o}}

\newcommand{\RR}{{\mathbb{R}}}

\def\al{\alpha}

\def\si{\sigma}
\def\om{\omega}

\def\Om{\Omega}

\newcommand{\beq}{\begin{equation}}
\newcommand{\beqa}{\begin{eqnarray}}
\newcommand{\eeqa}{\end{eqnarray}}

\begin{document}

\setlength{\unitlength}{1mm}

\thispagestyle{empty}
 \vspace*{2cm}

\begin{center}
{\bf \LARGE Black Holes as Lumps of Fluid}\\

\vspace*{1.5cm}

{\bf Marco M.~Caldarelli,}$^{1,2}\,$ {\bf \'Oscar J.~C.~Dias,}$^{1,3}\,$
{\bf Roberto Emparan,}$^{1,4}\,$ {\bf Dietmar Klemm,}$^5\,$

\vspace*{0.5cm}

{\it $^1\,$Departament de F{\'\i}sica Fonamental, Universitat de
Barcelona, \\
Marti i Franqu{\`e}s 1,
E-08028 Barcelona}\\[.3em]
{\it $^2\,$Instituut voor Theoretische Fysica, Katholieke Universiteit Leuven, \\
Celestijnenlaan 200D B-3001 Leuven, Belgium}\\[.3em]
{\it $^3\,$Dept. de F\'{\i}sica e Centro de F\'{\i}sica do Porto,
Faculdade de Ci\^encias da Universidade do Porto, Rua do Campo
Alegre 687, 4169 - 007 Porto, Portugal}\\[.3em]
{\it $^4\,$Instituci\'o Catalana de Recerca i Estudis Avan\c{c}ats (ICREA),
\\
Passeig Llu\'{\i}s Companys 23, E-08010 Barcelona, Spain}\\[.3em]
{\it $^5\,$Dipartimento di Fisica dell'Universit\`a di Milano, \\
Via Celoria 16, I-20133 Milano and \\
INFN, Sezione di Milano, Via Celoria 16, I-20133 Milano.}\\[.3em]

\vspace*{0.3cm} {\tt caldarelli@ub.edu, oscar.dias@fc.up.pt,
emparan@ub.edu, dietmar.klemm@mi.infn.it}

\vspace*{1cm}

\vspace{.8cm} {\bf ABSTRACT}
\end{center}

The old suggestive observation that black holes often resemble lumps of
fluid has recently been taken beyond the level of an analogy to a
precise duality. We investigate aspects of this duality, and in
particular clarify the relation between area minimization of the fluid
vs.\ area maximization of the black hole horizon, and the connection
between surface tension and curvature of the fluid, and surface gravity
of the black hole. We also argue that the Rayleigh-Plateau instability
in a fluid tube is the holographic dual of the Gregory-Laflamme
instability of a black string. Associated with this fluid instability
there is a rich variety of phases of fluid solutions that we study in
detail, including in particular the effects of rotation. We compare them
against the known results for asymptotically flat black holes finding
remarkable agreement. Furthermore, we use our fluid results to discuss
the unknown features of the gravitational system. Finally, we make some
observations that suggest that asymptotically flat black holes may admit
a fluid description in the limit of large number of dimensions.

\noindent


\vfill \setcounter{page}{0} \setcounter{footnote}{0}
\newpage

\tableofcontents

\newpage

\setcounter{equation}{0}\section{Introduction}

The tantalizing similarities between, on the one hand, the physics of
black holes, and on the other hand, the properties of soap bubbles and
fluid lumps, have been observed from old and indeed motivated approaches
like the membrane paradigm. A black hole is bounded by a smooth horizon
with uniform surface gravity, in analogy with the surface of a fluid
droplet which, under the effect of surface tension, assumes a smooth
shape with constant mean curvature. The negative specific heat of a
black hole, and its concomitant Hawking radiation, bear a resemblance to
the evaporation of a fluid droplet. More recently,
refs.~\cite{Cardoso:2006ks,Cardoso:2007ka,Cardoso:2006sj} pointed out a
striking similarity between the Rayleigh-Plateau instability of a
column of fluid and the Gregory-Laflamme instability of black strings
\cite{Gregory:1993vy,Gregory:1994bj}.

However, when one tries to take these observations beyond the level of
mere analogies, a number of apparent difficulties arise. A soap bubble,
or a lump of fluid ---at least a static one--- strives to minimize its
surface area, whereas a black hole tends to maximize the area of its
horizon. It is also difficult to make precise in which sense the
surface gravity relates to the surface tension of the fluid, or whether
it should be related instead to the mean curvature of the fluid surface.
Underlying these difficulties is the question of whether there is any
regime in which the gravitational dynamics of black hole horizons can be
precisely mapped into the dynamics of fluids.

The issue can be decisively settled within the context of the AdS/CFT
correspondence, which turns the analogy into a precise duality at least
for a certain class of black holes in Anti-de~Sitter space.
Refs.~\cite{Bhattacharyya:2007vs,Bhattacharyya:2008jc} have shown that
large black holes in AdS gravity can be mapped to solutions of the
Navier-Stokes equations of a conformal fluid. In this framework, large
black holes in a spacetime asymptotic to AdS with boundary $\cal B$
correspond to fluid configurations filling the spacetime $\cal B$.
However, the physics of these black holes differs significantly from the
physics of, say, the Schwarzschild black hole in asymptotically flat
space. For instance, large AdS black holes have positive specific heat,
and thus do not disappear via Hawking emission. In fact, the analogies
mentioned above seem to lose some of their motivation since the fluid
duals to these black holes do not have a bounding surface. Also, black
strings exist in AdS$_d$, with horizon topology $\RR\times S^{d-2}$
extended in a direction along an asymptotic boundary of topology
$\RR_t\times \RR\times S^{d-2}$
\cite{Copsey:2006br,Mann:2006yi,Bernamonti:2007bu}, but when these black
strings are large they do not suffer from the Gregory-Laflamme
instability \cite{Brihaye:2007ju}. In dual terms, they correspond to a
fluid filling the boundary. This fluid is locally stable under plasma
oscillations (\ie sound waves propagating in the fluid) so in the
long-wavelength approximation these configurations are stable. Thus,
these set-ups do not seem adequate to make precise the connection
between black holes and fluid \textit{lumps}, and in particular the
instabilities that may beset them. We need the dual fluid to admit a
bounding surface.

Such bounded fluid configurations arise naturally for gauge theories
with a deconfinement transition of first order. At temperatures just
above the deconfinement transition, lumps of deconfined plasma,
described at long wavelengths as fluids, can exist separated by a domain
wall from the confining vacuum \cite{Aharony:2005bm}. A gravitational
dual of a theory with this behavior is provided by Scherk-Schwarz (SS)
compactified AdS$_{d+2}$ gravity, whose solutions asymptote to
$\mathbb{M}^{d-1}\times \calS^1$, with $\calS^1$ standing for the
distinguished SS circle. Under the approximation of a very thin domain
wall, there are solutions for finite lumps of deconfined plasma whose
dual gravitational solution must correspond to black objects localized
near the infrared end of SS-AdS. These solutions are perfectly regular,
provided that the SS circle shrinks to zero size in the bulk, in
correspondence with the domain wall. This implies that the topology of
the corresponding event horizon is given by the fibration of the SS
circle $\calS^1$ over the plasma lump geometry, with the circle
shrinking to a point on the boundary. Plasma balls and plasma rings
correspond, respectively, to spherical black holes and to black rings in
the bulk. These were studied in \cite{Lahiri:2007ae} with an emphasis on
the AdS$_5$ case, and the full phase diagram for balls and rings in
AdS$_6$ was obtained more recently in \cite{Bhardwaj:2008if}. Crucially,
the horizon of the black hole is not mapped to the boundary of the
fluid, but to its entire bulk. Then the apparent contradiction between
minimal fluid surface area and maximal horizon area can be resolved.

Thus, this seems to be the correct framework to address the
correspondence between black holes and lumps of fluid. One aim of this
paper is to clarify the puzzles discussed above pertaining the
variational principles that determine the equilibrium configurations of
a fluid and of a black hole, as well as the relation between physical
magnitudes in both cases.

Having dealt with this, we attempt to uncover in the physics of these
plasma lumps some of the characteristic phenomena that
higher-dimensional black holes exhibit (see \eg
\cite{Emparan:2008eg,Kol:2004ww,Obers:2008pj}). In particular, we focus
on plasma tubes, which are dual to black strings. As mentioned above,
fluid analogues of the Gregory-Laflamme instability and of
non-uniform black strings \cite{Gubser:2001ac,Wiseman:2002zc} that
branch off at its threshold have been investigated recently
\cite{Cardoso:2006ks,Cardoso:2007ka,Cardoso:2006sj,Miyamoto:2008rd,
Miyamoto:2008uf,Dias:2007hg}. Here we reproduce and extend these results
to include rotation, which uncovers novel phases of non-uniform tubes.
But now we can regard these results as more than an analogy: following
\cite{Bhattacharyya:2008jc} they are actual solutions to a controlled
approximation to the Einstein equations in SS-AdS and thus describe
gravitational dynamics of black holes and black strings localized near
the infrared in the SS-AdS spacetime\footnote{Strictly speaking, the
analysis of \cite{Bhattacharyya:2008jc} refers to black branes dual to
fluids that fill the spacetime they live in. We can reasonably expect
that the extension to the situation of interest here should only be
technically more complicated due to the absence of analytic solutions
for plasma balls.}.

It turns out that these results show remarkable agreement with those
found for black holes and black strings in \textit{asymptotically flat}
space. This prompts the question of whether the latter should admit, in
some limit, a precise description in terms of fluid dynamics. After all,
even if the equivalence between AdS black holes and fluids is very
satisfying and promising, it does not yet explain the observations that
suggested an analogy between asymptotically flat black holes and fluids
in the first place. While we do not have a complete answer to this, we
will make some observations that suggest that asymptotically flat black
holes may be accurately described by fluid equations in the limit of
very large number of dimensions.

The plan of the rest of the paper is the following:
Section~\ref{sec:ConsTuv} reviews the basics of the relativistic
hydrodynamics of plasma lumps, and provides the dual identification
between temperature of the black hole and parameters at the surface of
the fluid. Section~\ref{sec:Variation} discusses the equivalence between
several variational principles for fluids. This allows to understand how
maximization of the entropy of a black hole can be equivalent to
minimization of the surface area of a fluid. In section~\ref{sec:EqConf}
we explore static and rotating configurations of plasmas in equilibrium
in a space with one (large) compact direction, with a particular
emphasis on tubes extended along this direction. In
section~\ref{sec:RP}, a detailed analysis of the Rayleigh-Plateau
instability for these is performed. We obtain the dispersion relations
for the unstable modes, both for static (Fig.~\ref{fig:GLstatic}) and
spinning tubes (Fig.~\ref{fig:GLrotate}). After these analyses, we
discuss in section~\ref{sec:RegimeValidity} their range of validity, and
in section~\ref{sec:RPGL} how they compare to the phases and stability
of black strings in {\it vacuum} gravity. Section~\ref{sec:revisit}
revisits the issues posed at the beginning of the paper and suggests
that vacuum black holes in the limit of large number of dimensions may
admit a fluid description.

\bigskip

{\it Note added:} during the final stage of this project we have been
informed of the work \cite{mandm}, which overlaps with some of our results.

\setcounter{equation}{0}
 \section{Hydrodynamic description of deconfined plasma lumps\label{sec:ConsTuv}}

\subsection{Navier-Stokes and Young-Laplace equations}

Large black holes in AdS correspond to deconfined gluon  plasma in
the boundary CFT. In the long-wavelength approximation, this plasma
behaves as a fluid, whose properties can be deduced from its
gravitational bulk description. In this approximation, one can
perform a derivative expansion, whose leading order is given by a
perfect fluid, while next orders introduce dissipative and diffusive
phenomena. More precisely, it has been shown that the
long-wavelength sector of pure AdS gravity is described by
Navier-Stokes equations on the boundary, and the stress tensor of
the corresponding fluid has been computed up to second order in the
derivative expansion in five
\cite{Bhattacharyya:2008jc,Bhattacharyya:2008ji} and four dimensions
\cite{VanRaamsdonk:2008fp}.

We are interested in the case where the boundary theory admits a
confining phase, which close to the deconfinement temperature is
separated from the deconfined plasma by a thin domain wall
\cite{Aharony:2005bm}. We assume that our plasma lumps are large
enough to neglect the thickness of the wall. The analysis of this
subsection follows largely \cite{Lahiri:2007ae}. We shall neglect
subleading dissipation and diffusion contributions\footnote{In
particular, for stationary equilibrium solutions, these
contributions vanish. We will comment on the effects of higher order
corrections on the analysis of the perturbations near the end of the
article.}. In this approximation, the deconfined plasma fluid
behaves as a perfect fluid, and the leading order stress tensor is
given by the sum of a perfect fluid part and a boundary surface
contribution, describing the capillarity of the fluid,
\begin{eqnarray}
&&T^{\mu\nu}=T^{\mu\nu}_{\rm perf}+T^{\mu\nu}_{\rm bdry}\,,\nonumber\\
&&T^{\mu\nu}_{\rm perf}=\left[\lp\rho+P\rp u^\mu u^\nu+P
g^{\mu\nu}\right]\Theta(-f) \,,\qquad T^{\mu\nu}_{\rm bdry}= -\sigma
h^{\mu\nu}|\partial f|\,\delta(f)\,.\label{GenLumpTuv}
\end{eqnarray}
Here, $u^{\mu}$ is the fluid velocity, $\rho$, $P$ and $\sigma$ are
its density, pressure and surface tension respectively. The fluid
boundary is defined by $f(x^{\mu})=0$, has unit normal
$n_{\mu}=\partial_{\mu} f/|\partial f|$ pointing towards the confining
phase, and
$h^{\mu\nu}=g^{\mu\nu}-n^\mu n^\nu$ is the projector onto the
boundary. The velocity field $u^\mu$ is subject to the boundary condition
\eq
 u^\mu n_{\mu} = 0 \,, \label{constraintYL}
\eeq
requiring that the fluid velocity must be orthogonal to the boundary normal
or else the fluid would not be confined inside the boundary but flow through it.
Finally, $\Theta(-f)$ is the Heaviside function
($\Theta(-f)=1$ inside the fluid and zero everywhere
else).

The equations describing the dynamics of the fluid follow  from the
conservation of the energy momentum tensor (\ref{GenLumpTuv}). The
volume and boundary contributions must vanish independently; the
former, projected along the fluid velocity $u_{\nu}$ yields the
relativistic continuity equation,
\eq u^{\mu}\nabla_{\mu}\rho + (\rho+P) \nabla_{\mu} u^\mu = 0\,,
\label{continuity} \eeq
(where $\nabla_{\mu} u^\mu=\theta$ is the expansion of the fluid)
that we use to reduce the remaining equations to the relativistic
Navier-Stokes equation,
\eq
(\rho+P) u^\mu \nabla_{\mu} u^\nu = -\lp g^{\mu\nu}+u^\mu u^\nu \rp
\nabla_{\mu} P \,. \label{NavierS}
\eeq
Here $ u^\mu \nabla_{\mu} u^\nu=a^{\nu}$ is the acceleration of the
fluid and $g^{\mu\nu}+u^\mu u^\nu$ is the projector onto
the subspace orthogonal to $u^{\mu}$.

Now consider the boundary contribution to the conservation of the stress tensor,
\eq
\left[(\rho+P) u^\mu u^\nu+P g^{\mu\nu}\right] |\partial f| n_{\mu}
+ \sigma\nabla_{\mu}\lp h^{\mu\nu} |\partial
f|\rp=0\,.\label{consTuv2}
\eeq
Projecting this equation onto $n_{\nu}$, using (\ref{constraintYL}), and
integrating across the boundary we get
\eq
P_<-P_>=\sigma K\,, \qquad K\equiv h_{\mu}^{\:\:\nu}\nabla_{\nu}
n^{\mu}\,, \label{YoungLap}
 \eeq
where $K$ is the trace of the boundary's extrinsic curvature (twice
the mean curvature $H$ of the bounding surface), and $P_<-P_>$ is
the pressure jump when we cross the boundary from the exterior  into
the interior. Equation (\ref{YoungLap}) is the Young-Laplace
equation that describes the capillary pressure difference $\Delta P$
sustained across the interface boundary due to surface tension
$\sigma$ \cite{young}.

\subsection{Stationary plasma configurations\label{sec:StationaryPlasma}}

The Young-Laplace equation we found in the last subsection expresses
the balance of forces on the plasma boundary and holds in the most
general dynamical situation. We will turn now to the study of
stationary plasma configurations in hydrodynamical and
thermodynamical equilibrium.

To begin with, we need to characterize stationary configurations of
plasma. Let us assume that the background spacetime is stationary, with
stationarity timelike Killing vector $\xi$, and a set of commuting,
linearly independent spacelike Killing vectors $\chi_I$, corresponding
to the isometries of the background. Of the latter, the subset that
commutes with the velocity field will also be symmetries of the fluid,
${\cal L}_{\chi_I}u=0$. The associated conserved charges will be the
energy $E$ and the linear/angular momenta $J_I$ of the plasma associated
to fluid symmetries.

Let $\calP$
be the spatial region filled by the
plasma, $\p\calP$ its boundary, and $u^\mu$ the velocity field of
the fluid. Stationarity, for an isolated system, requires that there
be no dissipation, and therefore the bulk and shear viscosity terms
of the stress tensor must vanish. This condition is met if and only
if the expansion $\theta$ and the shear $\si^{\mu\nu}$ of the fluid
both vanish. Under these assumptions, one can show that there exists
a function $\al$ such that $\al u^\mu$ is a Killing vector.

Indeed, for a shearless, expansion-free velocity field,
\eq
\nabla^\mu u^\nu=\omega^{\mu\nu}-u^{\mu}a^{\nu}, \label{nablau}
\eeq
where $\omega^{\mu\nu}$ is the vorticity of the fluid and
$a^\mu$
its acceleration. Moreover, a fluid with local entropy density $s$
and local temperature ${\cal T}$ has to satisfy the Euler relation,
\eq \rho+P ={\cal T} s\,. \label{Euler}
\eeq
Differentiating this
equation and using the first law of thermodynamics, we get the
Gibbs-Duhem relation $dP=sd{\cal T}$. Hence, since for a stationary
configuration $u^\mu\nabla_\mu P=0$, the Navier-Stokes equations can
be rewritten as
\eq a_\mu=-(\rho+P)^{-1}\nabla_\mu
P=-\nabla_\mu\ln\T, \label{heat}
\eeq
where, in the last step, we
used both the Euler and the Gibbs-Duhem relations. Note that this
equation expresses simply the fact that the heat flux vanishes
\eq
q^\mu = -\kappa (g^{\mu\nu}+u^\mu u^\nu) (\nabla_\nu{\cal T} +
a_\nu{\cal T})=0
\eeq
or, in other words, stationary plasma
configurations are both at hydrodynamical and thermal equilibrium.
Substituting equation (\ref{heat}) in (\ref{nablau}) it follows
that
\eq \nabla_{(\mu}(\al u_{\nu)})=\al
u_{(\mu}\nabla_{\nu)}\ln(\al\T).
\eeq
Hence, if we choose the
function $\al$ to be inversely proportional to the local temperature
of the plasma, $\al=T/\T$ with $T$ an integration constant, the
vector field $\al u^\mu$ solves the Killing equations, and must
therefore be a linear combination of the Killing vectors $\xi$ and
$\chi_I$\footnote{Actually, it could depend linearly on Killing
vectors not commuting with $\chi_I$. However, one can always choose
the set of Killing vectors $\chi_I$ to be adapted to the motion of
the fluid under consideration.}. Hence, stationary configurations
have a velocity field given by
\eq u=\frac\T T\lp
\xi-\Omega_I\chi_I\rp\,. \label{velocityfield}
\eeq
The Killing vectors appearing here are a subset, although not
necessarily the full set (since there may be rotational symmetries with
no rotation velocity of the fluid in their direction), of the abelian
symmetries of the fluid, which themselves are a subset of the symmetries
of the background. The most general solution is determined by the
constant parameters $T$ and
$\Om_I$, since the local temperature $\T$ is then fixed by the
relation $u^2=-1$ to be
\eq \T = \gamma\, T\,, \label{localtemp} \eeq
where
\eq\label{gammxiom}
\gamma^{-1}=\sqrt{-(\xi-\Om_I\chi_I)^2}
\eeq
is the redshift factor relating measurements
done in the laboratory and comoving frames, and the constant $T$ is
the equilibrium plasma temperature. It follows from the general form
(\ref{velocityfield}) of $u^\mu$ that a stationary equilibrium fluid
configuration has to be in rigid roto-translational motion.

Combining the Euler relation (\ref{Euler}) and the Young-Laplace
equation (\ref{YoungLap}), we can relate the temperature parameter $T$
of the fluid (which, for a non-boosted lump in an ultra-static
background spacetime with $\xi^2=-1$, is the temperature of the fluid at
the axis of rotation) to a combination of several magnitudes at the
fluid surface,
\eq\label{TsigmaK}
T=\frac{\sigma K +\rho}{\gamma s}\,.
\eeq
In the duality to a black hole, $T$ corresponds
to the Hawking temperature of the horizon. We see that $T$ is \textit{not}
simply proportional to the surface
tension or to the mean curvature, although it grows linearly with either of
them. Note also that for a static fluid $K$ will be constant over the
surface, but in a stationary configuration the curvature $K$ will adjust
itself to variations of the fluid velocity near the boundary. Let us
also mention that the angular velocities of the
black hole horizon are identified with the $\Omega_I$.

So far we have not specified the equation of state of the fluid. We
are interested in the $d$-dimensional (non-conformal) plasma
describing the hydrodynamic limit of the ($d+1$)-dimensional CFT dual
to Scherk-Schwarz reduced AdS$_{d+2}$ gravity, whose equation of
state in $d=n+3$ dimensions reads (see \cite{Lahiri:2007ae})
\begin{equation}
\rho+P=\frac{n+4}{n+3}\lp \rho-\rho_0\rp \,,
\label{ConfEqState}
\end{equation}
and
\begin{equation}
s=(n+4)\alpha^{1/(n+4)}
\left(\frac{\rho-\rho_0}{n+3}\right)^{\frac{n+3}{n+4}}\,,\qquad
{\cal T}=\left(\frac{\rho-\rho_0}{(n+3)
\alpha}\right)^{\frac{1}{n+4}}\,,
 \end{equation}
with $\rho_0$ and $\alpha$  constants. This equation of state is
normalized such that the vacuum pressure outside the plasma
vanishes. When $n=0$, if we consider the theory that is dual to type II
strings in SS-AdS$_5\times S^5$, we have
\eq \al=\frac{\pi^2N^2}{8T_c}\,,\qquad
\rho_0=\frac{\pi^2N^2T_c^3}8\,.
 \eeq
  The surface tension of the
domain wall has been computed numerically in \cite{Aharony:2005bm}
for $n=0$ and $n=1$. In units of $\rho_0/T_c$ they obtain
 \eq
  \si_{n=0}=2.0 \frac{\rho_0}{T_c}\,,\qquad {\rm and}\qquad \si_{n=1}=1.7
\frac{\rho_0}{T_c}\,.  \label{numsigma}\eeq

For a plasma in equilibrium, it follows from (\ref{localtemp}) and
(\ref{ConfEqState}) that the density and pressure satisfy the
relations
\eq \rho=\rho_{*} \gamma^{n+4} +\rho_0\,, \qquad
P=\frac{\rho_{*}}{n+3}\,\gamma^{n+4}-\rho_0\,,\label{DpEquil}
 \eeq
where
\eq
\rho_{*}=(n+3)\alpha T^{n+4}
\eeq
is the difference $\rho-\rho_0$ between the plasma energy density and the
energy density of the confining vacuum at points where $\gamma=1$. We see that all
the dependence of the density on the position is entirely given by the
$\gamma$ factors.

\setcounter{equation}{0}
\section{Variational principles for equilibrium plasma configurations
\label{sec:Variation}}

In this section we will show that the Young-Laplace equation
(\ref{YoungLap}) can also be obtained using two equivalent variational
principles, namely through a maximization of entropy and/or through a
minimization of potential energy. As an easy byproduct we will see that
these are also equivalent to minimization of the surface area for static
fluids. Our formulation will be relativistic, fully covariant, and valid
for a large class of stationary background spacetimes. While the
relation between these two variational principles is probably
well-known, we were not able to find it in the literature and we
surmise that the relativistic covariant generalization that we present
is original.

\subsection{Relativistic soap bubbles and second law of thermodynamics}

Equilibrium plasma configurations must have constant temperature and
are expected to be those that extremize the entropy of the plasma
while keeping its energy and momenta fixed. We will use this
variational principle to give an alternative derivation the
Young-Laplace equation, shedding a new light on the relation between
black holes and their membrane analogues.

As explained in the previous section, equilibrium configurations on
a stationary background spacetime are in rigid motion, and have a
velocity field $u^\mu$ of the form (\ref{velocityfield}), determined
by the temperature $T$ of the plasma and the boost parameters
$\Omega_I$. Again $\xi^\mu$ is the stationarity Killing vector
and we choose a time coordinate $t$ such that $\xi=\p_t$. This
allows us to foliate the spacetime into constant $t$ hypersufaces
$\Sigma_t$, that we use to define the conserved charges. We call
$k^\mu$ the unit normal vector to these hypersurfaces, and we make the
additional hypothesis that the Killing vectors $\chi_I$ generate
isometries of $\Sigma_t$, \ie $k\cdot\chi_I=0$.
Then, given any Killing vector $\psi^\mu$, one can define the associated
conserved charges
\eq {\mathcal Q}[\psi]=\int_{\Sigma_t}\!\!dv
\;T_{\mu\nu}k^{\mu}\psi^\nu, \eeq
where $dv$ is the induced volume measure on $\Sigma_t$.

To perform the maximization of the entropy at constant energy and
momenta, we define the action
\eq I[\calP]=S[\calP]-\beta
E[\calP]+\tilde\omega_I J_I[\calP]\,,\label{Ientropy} \eeq
where $\beta$ and
$\tilde\om_I$ are the Lagrange multipliers associated to our constraints.
The total entropy of the fluid is the conserved charge associated
to the entropy density current $su^\mu$, \eq S[\calP]=-\intP (k\cdot
u) s\,dv\,,\label{Var:Entropy} \eeq while the energy and momenta are
given by
\begin{eqnarray}
&& E[\calP]=\intP\left[(\rho+P)(\xi\cdot u)(k\cdot u)+\lp k\cdot\xi\rp P\right]
    dv-\sigma \int_{\Sigma_t} k^\mu\xi^\nu
h_{\mu\nu}|\p f|\delta(f)\,dv   \,,\nonumber\\
&& J_I[\calP]=\intP\lp\rho+P\rp(k\cdot u)(\chi_I\cdot u)\,dv
-\sigma\int_{\Sigma_t} h_{\mu\nu}k^\mu\chi^\nu_I|\p f|\delta(f)dv
\,.
 \label{Var:charges}
\end{eqnarray}
In addition to the bulk terms, both the energy $E$ and the momenta $J_I$
have a surface contribution proportional to the surface tension. The
boundary term in $E[\calP]$ corresponds to the surface tension potential
energy, and reduces to $\sigma$ times the area of the boundary of the
fluid lump for a configuration respecting the spacetime symmetries in an
ultra-static background. The boundary term in $J_I[\calP]$ is
present only if $\chi_I\cdot n\neq 0$ so the fluid boundary is not
invariant under the action of $\chi_I$, for instance when the surface is
not axisymmetric.

Taking the Lagrange
multipliers to be given by
\eq
 \beta=1/T\,,\qquad
 \tilde\omega_I=\beta\Omega_I\,,
 \label{multipliers}
\eeq
making use of the form (\ref{velocityfield}) for $u$ and eliminating the
entropy density $s$ by using the Euler relation, the action simplifies to
\eq
I[\calP]=-\beta\intP
(k\cdot\xi) P\,dv
+\beta\sigma\int_{\Sigma_t}(k\cdot\xi)|\p f|\delta(f)\,dv.
\label{simpaction}
\eeq

To extremize this action, we need to find its variation under a
deformation of the region $\calP$ occupied by the fluid. Notice that,
since by hypothesis the configuration is stationary, the action
is time independent, and its extrema are not changed if we integrate
it over the time $t$. Then, noting that the spacetime volume element
is
\eq
dV=-(\xi\cdot k)\,dv\,dt\,,
\eeq
it follows
that the quantity we have to extremize assumes the form
\eq
\int I[\calP]\,dt = \beta\int_{\cal M}P\Theta(-f)\,dV
-\beta\sigma\int_{\cal M}|\p f|\delta(f)\,dV\,,
\label{spacetimeaction}\eeq
where the integrals extend to the full spacetime $\cal M$, and the
resulting functional is manifestly covariant. Also, note that the last
integral is the area of the region of spacetime (world-volume) spanned
by the boundary of the fluid. To extremize (\ref{spacetimeaction}) under
deformations of $\calP$, we compute its variation under a change $\delta
f$ of the function $f$ and integrate by parts the last term,
\eq
\delta_f\int I[\calP]\,dt = \beta\int_{\cal M}
\lp P-\sigma K\rp\delta f\,\delta(f)\,dV\,.
\label{variation}\eeq
We used here the well-known fact that the variation of
the area of an hypersurface is precisely given by twice the mean
curvature, or its extrinsic curvature $K$. Therefore, requiring that
(\ref{variation}) vanishes for any deformation $\delta f$ of the
boundary, we obtain that the variational principle boils down to
\eq \sigma K=P\,, \label{YoungLap2} \eeq
where $P$ is the pressure jump at the boundary of the fluid. This,
for non-vanishing external pressure, is again the Young-Laplace
equation (\ref{YoungLap}) that we derived in the last section using
the conservation of the energy-momentum tensor. This time we found
that it also follows from requiring extremization of plasma entropy
for fixed energy and momenta. Note that the derivation assumes only
the stationarity of the configuration and of the background
geometry, and is independent of the fluid equation of state.
Besides, the derivation does not assume any condition on the shape
of $\calP$, in particular it covers non-axisymmetric cases where the
geometry of the fluid configuration does not share the symmetries of
the fluid motion, as in the case of the two-lobed figures of equilibrium.

\subsection{Equivalent variational principles for spinning lumps}

A universal behavior of fluids is that they always pick boundary
configurations that reduce their potential energy for a fixed volume.
For static solutions, this implies that the area of the fluid surface is
minimized. For stationary solutions, the potential energy not only has a
surface tension term but also a centrifugal contribution. We want to
verify the equivalence between this potential energy minimization (hence
a mechanical formulation) problem and the entropy maximization problem
(a thermodynamic formulation) studied in the previous subsection.

To find an energetic formulation of the variational principle, we
first observe that, using \eqref{velocityfield}, the total action
(\ref{Ientropy})
assumes the form (\ref{simpaction}) in which the second term in
the right hand side has the interpretation of surface
potential energy, while the pressure integral has the interpretation
of centrifugal potential energy, since the radial variation of
pressure measures the centrifugal force per unit volume of fluid.

Therefore, we define a new action
\eq \widehat{I}[\calP]
=U_{\sigma}[\calP]+U_{\rm cf}[\calP]-\eta
V[\calP]\,,\label{Ipotential} \eeq
where $U_{\sigma}[\calP]= \sigma \A[\dP]$ is the
surface tension potential energy associated to the area
\eq \A[\dP]=-\int_{\Sigma_t} (k\cdot\xi)
|\p f|\delta(f)\,dv, \eeq
the centrifugal potential energy, obtained by integrating the
pressure over the plasma volume is \eq U_{\rm
cf}[\calP]=\intP\,(k\cdot\xi)(P-P_>)\,dv \,,\label{CentrifPot} \eeq
with $P_>$ an integration constant, and finally the volume
$V[\calP]$ reads \eq V[\calP]=-\intP (k\cdot\xi)\,dv\,, \eeq and is
kept fixed as we perform the variation through the Lagrange
multiplier $\eta$. Note that, in an ultra-static background,
$\A[\dP]$ and $V[\calP]$ reduce to the usual definitions of boundary
area and volume of the plasma, respectively. Again, one can then
easily verify that
the Euler-Lagrange equation associated with (\ref{Ipotential})
yields the Young-Laplace equation (\ref{YoungLap}), and the Lagrange
multiplier takes the value $\eta=P_>$, where $P_>$ is the exterior
vacuum pressure.

The two variational approaches, maximization of entropy and
minimization of area, are equivalent because the actions
(\ref{Ientropy}) and (\ref{Ipotential}) are proportional to each
other. Indeed, starting with the action (\ref{Ientropy}), successive
use of (\ref{Var:charges}), (\ref{multipliers}) and (\ref{Euler})
yields the following relation between actions (\ref{Ientropy}) and
(\ref{Ipotential}),
\eq I[\calP]=-\beta \widehat{I}[\calP] \,.\label{relationI's}
 \eeq
Since the two actions are the same up to a negative constant, minimizing
the potential energy for fixed volume, $\delta
\widehat{I}[\calP]=0$, is naturally equivalent to maximizing the
entropy for fixed conserved charges, $\delta I[\calP]=0$.

Finally note that in static configurations, $\Omega_I=0$,
the
pressure is constant inside the fluid and we can write
\eq I[\calP]=-\beta
\widehat{I}[\calP]=-\beta\sigma\lp\A[\dP]-KV[\calP]\rp
\,,\label{nonRotI's}
 \eeq
and when this action is extremized the Lagrange multiplier $K$ is
required to be twice the mean curvature.
Therefore, non-rotating equilibrium fluids have boundary shapes that
minimize the surface area for fixed volume and their boundaries have
constant mean curvature, a well known property of fluids.

\bigskip

Black holes satisfy the variational principle that their entropy is
extremized for fixed energy and angular momenta --- this is essentially
the first law. In the duality between SS-AdS black holes and fluid
lumps, the entropy, energy and spins are identified on both sides, while
the temperature is mapped according to eqs.~\eqref{TsigmaK} and
$\Omega_I$ are the angular velocities. Then our analysis shows that
maximization of the black hole horizon area is equivalent, for static
configurations, to minimization of the fluid surface area. When rotation
is present, the connection is less simple geometrically, but still
easily expressed as an extremization of a functional.

Thus, in this context the proposal in \cite{Cardoso:2006ks} to model
black strings with non-gravitating cylinders of fluids can be made
exact. There are, however, some differences between the analysis in
refs.~\cite{Cardoso:2006ks,Miyamoto:2008rd} and ours: theirs was a study
of non-relativistic fluids, whereas our fluids are relativistic. For
static fluids the differences are small\footnote{Namely, $\rho+\calP
\sim \rho$ in the non-relativistic limit and the extrinsic curvature has
time derivative terms that disappear in the non-relativistic limit.
These time derivative terms are however vanishing when dealing with
stationary solutions.}, but for non-static or out of equilibrium
situations the relativistic corrections become crucial. But the more
important differences are that, first, we have a definite equation of
state for the fluid and a specification of the value of the surface
tension. This allows us to compute thermodynamic quantities of the fluid
that are identified with those for the black hole, something that was
impossible in \cite{Cardoso:2006ks,Miyamoto:2008rd}. Second, the number
of dimensions for the fluid and the black hole are different: a fluid in
$d$ spacetime dimensions is mapped onto a black hole in $d+2$
dimensions. In particular, the entire volume of the fluid can be
regarded as mapped onto the black hole horizon.

\setcounter{equation}{0}
\section{Static and rotating equilibrium plasma configurations \label{sec:EqConf}}

In this section we specialize to the study of stationary axisymmetric
rigidly rotating plasma configurations in a flat spacetime and discuss
the corresponding phase diagrams of solutions. These also represent the
phase diagrams for the dual black objects. We start with a general
description of the system we want to study, and then we obtain the
differential equation for the shape of equilibrium plasmas. We find the
static plasma configurations in any dimension $n$, and the rotating
plasma configurations for $n=1$ and discuss the resulting phases.

\subsection{General formulation of the stationary and axisymmetric case\label{sec:EquilCond}}

The shape of stationary and axisymmetric plasma configurations is
determined  by the Young-Laplace equation. The background geometry
is $d=n+3$ dimensional flat spacetime $\mathbb{R}_t\times
\mathbb{R}^{n+1}\times S^1$ (with $n\geq1$), and we choose to work
in coordinates such that the metric reads
\eq
ds^2=-dt^2+dr^2+r^2(d\theta^2+\sin^2\theta\,d\phi^2+\cos^2\theta\,d\Omega^2_{n-2})+dz^2\,,
\label{metric} \eeq
where $\theta\in[0,\pi/2]$, $\phi\in[0,2\pi)$ and $d\Omega^2_{n-2}$
is the metric of the unit $(n-2)$-sphere. For $n=1$ all the results can
be reproduced by setting $\theta=\pi/2$.

We consider rigidly rotating configurations with fluid velocity
\eq
 u^{\mu}=\gamma \lp \delta^{\mu}{}_{t} +
\omega_{\phi}\delta^{\mu}{}_{\phi} \rp \,,\qquad
  \gamma=\left(1-g_{\phi\phi}\,\omega_{\phi}^2 \right)^{-1/2}\,,\label{velocity} \eeq
with constant angular velocity $\omega_{\phi}$. Note that a boost
$\omega_{z}$ as well as rotation in other planes (if allowed by the
dimension of the spacetime) could be easily added to our discussion.

These plasma configurations satisfy the equilibrium equation of
state (\ref{DpEquil}). Their entropy $S$ and conserved charges $E,
J$ are given, respectively, by (\ref{Var:Entropy}) and
(\ref{Var:charges}) with the relevant Killing vectors associated to
(\ref{metric}) being $\xi=\partial_t$ and $\chi_\phi=\partial_\phi$.
We will look for surfaces of
revolution invariant under the action of these vectors.

\subsection{Static plasma lumps \label{sec:StaticTube}}

Our discussion will be very succint since these configurations were
studied in \cite{Miyamoto:2008rd} and we have rederived their main
results. However, since we have an equation of state for the fluid we
are able to discuss the properties of the solutions in a different
manner: we display the phases in an entropy {\it vs} energy diagram.
These quantities map directly into the entropy and energy of the dual
black holes and black strings in SS-AdS. Therefore a direct comparison
is possible to the diagrams that have been obtained for the black hole
and black string phases in vacuum gravity.

For a profile of the form $r=R(z)$ the equations follow easily from the
Young-Laplace equation \eqref{YoungLap} which, as we have discussed
extensively, requires the mean curvature $K$ of the surface of a static
fluid to be constant.
The equilibrium equation of state (\ref{DpEquil}) fixes this constant to
\eq
K=\frac{\rho_*-(n+4)\rho_0}{(n+3)\sigma}\,.
\eeq

There are three families of solutions to this equation:

\begin{itemize}
\item{\it Uniform plasma tubes (UT):} These solutions have constant
radius $r=R_o$ along the circle, so $K=n/R_o$. These are essentially
configurations of a plasma ball in $d-1=n+2$ dimensions, times a
straight line $z$, and have pressure and density
\eq
P=n\frac{\sigma}{R_o}\,,\qquad
\rho=n(n+3)\frac{\sigma}{R_o}+(n+4)\rho_0\,.
\eeq

\item{\it Plasma balls (B):} These solutions are characterized by
 $R(z)=\sqrt{R_o^2-z^2}$, \ie they describe a $(n+1)$-dimensional
 sphere of radius $R_o$, and extrinsic curvature $K=(n+1)/R_o$, so
\eq
P=(n+1)\frac{\sigma}{R_o}\,,\qquad
\rho=(n+1)(n+3)\frac{\sigma}{R_o}+(n+4)\rho_0\,.
\eeq

\item{\it Non-uniform plasma tubes (NUT):} For $n\neq 0$, there is a third
family of static equilibrium solutions that describe non-uniform
plasma tubes. These configurations are found by solving numerically
a first integral of the equation for the mean curvature.
\end{itemize}
\begin{figure}[t]
\centerline{\includegraphics[width=.48\textwidth]{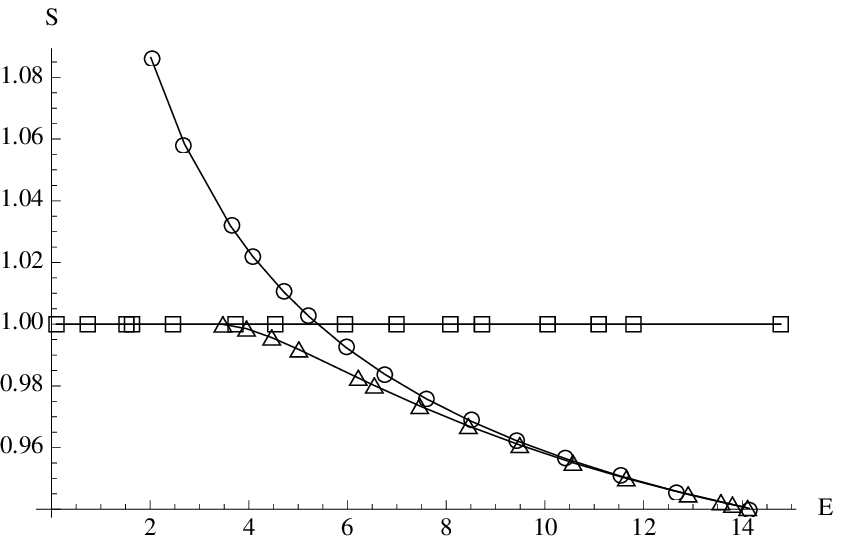}
\hfill\includegraphics[width=.48\textwidth]{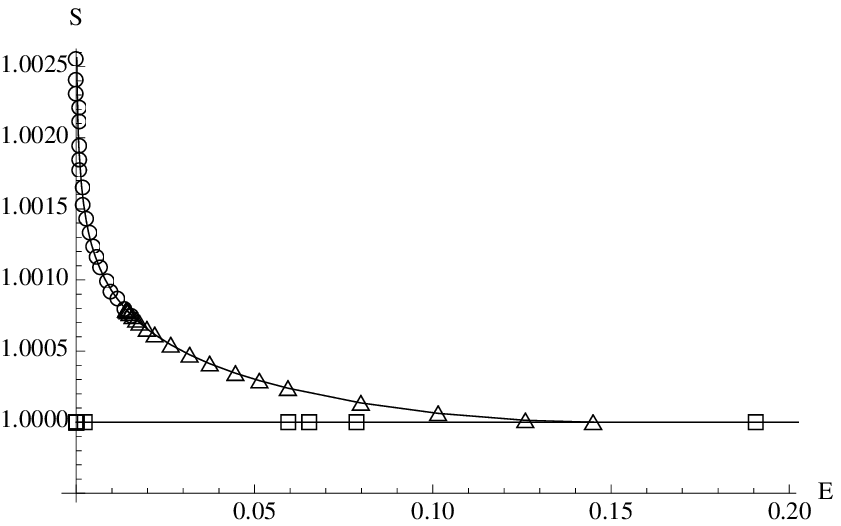} }
\caption{\small Entropy versus energy phase diagrams. The entropy is
normalized to be one for uniform plasma tubes. Uniform plasma tubes,
plasma balls and non-uniform plasma tubes correspond to squares,
circles and triangles respectively. The left diagram, plotted for
$n=1$, is representative for $n\leq7$, and shows that the preferred
configuration is the plasma ball for small energy, and the uniform
plasma tube for higher energy. The unstable plasma ball branch
merges with the non-uniform plasma tube branch in the merger point.
Above the critical dimension, for $n\geq8$ (the right diagram
corresponds to $n=12$), the non-uniform plasma tubes become stable
and become the preferred configuration for an intermediate range of
energies. The left figure should be compared with Fig.~3 of
\cite{Harmark:2005pp} for the phases of non-uniform black strings.
To plot these diagrams we used the values $\alpha=1$, $\rho_0=0$ and $\sigma=1$.
} \label{fig:staticphases}
\end{figure}

We analyzed these solutions numerically, and constructed the
corresponding phase diagrams in the $E-S$ plane. These are shown for
$n=1$ (left) and $n=12$ (right) in Fig.~\ref{fig:staticphases}. The
preferred configuration, at a given energy, is given by the one with
highest entropy. There is a critical dimension
$n_\star=8$ for which the qualitative behavior changes. For
$n\leq7$, the entropically favored configuration is the plasma ball
for small energy, which turns into a uniform plasma tube through a
first order phase transition as the energy is increased. The
non-uniform plasma tube branch has always a lower entropy
(Fig.~\ref{fig:staticphases}, left). However, for $n\geq8$, the
uniform plasma tubes benefit entropically from generating
non-uniformities in an intermediate range of energies
(Fig.~\ref{fig:staticphases}, right).

The case $n=0$ is special, since the fluid lives in a two-dimensional
space and the uniform plasma strips and plasma disks are the only
solutions with constant mean curvature. Static plasma strip solutions have
vanishing fluid pressure: since the section of the boundary at constant
$z$ consists of just two points, there is no extrinsic curvature of the
surface that must be balanced by pressure. This means that capillarity
does not exert any force in the direction orthogonal to the plasma
strip. This is consistent with the fact that the surface area does not
change when the tube radius changes. In this case, the configuration is
in a stable equilibrium, and no non-uniform plasma strip solution
exists.

In the next section, we shall obtain a better understanding of these
features in terms of the mechanical stability properties of the uniform
plasma tubes. Indeed, the existence of the non-uniform plasma tube
branch can be inferred from the fact that the uniform plasma tube
changes its stability properties for a critical energy, at which it
develops a static unstable mode. This critical unstable mode implies a
bifurcation point to a new branch of static solutions that describe the
non-uniform plasma tubes, and is associated to the Rayleigh-Plateau
instability. For $n=0$, the corresponding threshold mode has infinite
wavelength, so there cannot be any non-uniform branches connected to the
uniform strip.

Note also that one can have an arbitrary number of coexisting
disconnected plasma lumps. In the plasma fluid approximation, which is
valid as long as they are far apart compared to the mean free path of
the fluid, they do not feel each
other\footnote{For the dual black holes, this means separations larger than
the cosmological length scale $1/\sqrt{-\Lambda}$, and their mutual
gravitational interaction would involve massive
Kaluza-Klein modes in the vacuum phase, which is subleading.}. In
particular, other phases in which the fluid is distributed in multiple
plasma balls can be present, but they have subleading entropy with
respect to the single plasma ball in the phase diagram. Also, one can
construct new branches of non-uniform plasma tubes, known as {\em copy
solutions}, by unwrapping the original non-uniform plasma tube along the
compact $z$ direction, and changing its periodicity. These new branches
can be easily obtained through scaling arguments \cite{Horowitz:2002dc}.
Since they are subleading in entropy we will not consider them.

\subsection{Rotating plasma lumps \label{sec:RotTube}}

The problem of finding general equilibrium solutions in dimension $n$ is
considerably complicated by the presence of rotation, which breaks the
$SO(n+1)$ symmetry of the sections at constant $z$ to $SO(2)\times
SO(n-1)$ and introduces a dependence on the angle $\theta$ in the
profile of the lumps. Then one is forced to solve a partial differential
equation. However, this problem is not present when $n=1$, where the
transverse spheres are actually circles so there is the same symmetry as
in the static case and the equations reduce to ODEs. Therefore we
restrict our analysis of spinning lumps to $n=1$. Even if there may be
phenomena of higher-dimensional tubes that we miss by this restriction,
the case of $n=1$ still exhibits qualitatively new dynamics relative to
the static situation.

We represent the boundary of the spinning lumps in terms of a height function
$h(r)$,\footnote{For the uniform tubes, the simplest boundary
parametrization is instead $f(r)=r-R_o=0$ and the Young-Laplace
equation gives $\rho_c$ as a function of $\rho_0,R_o,\omega_\phi$.}
\eq\label{height}
f(r,z)=z-h(r)=0
\eeq
Then, use of the equilibrium equation of state (\ref{DpEquil}) for
the pressure and of the Young-Laplace equation (\ref{YoungLap})
yields the ODE for a stationary axisymmetric plasma configuration,
\eq \frac{d}{dr}\lp \frac{r h'}{\sqrt{1+h'^2}}\rp
+\frac{\rho_*}{4\sigma}\,r \lp
1-r^2\omega_\phi^2\rp^{-5/2}-\frac{\rho_0}{\sigma}\,r=0\,.
\label{rot:EqMotion}
 \eeq

Some solutions to these equations were already studied in detail in
\cite{Lahiri:2007ae,Bhardwaj:2008if}, namely, balls, pinched balls and
rings. Now they are constrained to fit along the $z$ direction, which we
are taking to be compact. Two other solutions, the rotating uniform tube
and the uniform `hollow tube'
can be readily constructed out of them by simply taking the rotating
plasma disk and the plasma annulus that appear in one less dimension
and translating them uniformly along
$z$.\footnote{Observe that this is not possible for pinched balls since
they do not exist for $n=0$.} To make contact with this previous work we
follow partially their notation and define the dimensionless
variables,
\eq \widetilde{\omega}_\phi=\frac{\sigma
\omega_\phi}{\rho_0}\,,\qquad v=\omega_\phi r\,,\qquad
H(v)=\omega_\phi h(r)\,,\qquad
k=\frac{1}{4}\frac{\rho_*}{\rho_0}\,,\qquad
\tilde{L}=\frac{\rho_0}{\sigma} L\,, \label{DlessVar} \eeq
where $L$ will be used to represent the length of the uniform tube
with radius $R_o$ and $\tilde{L}$ the corresponding dimensionless
quantity. Note also that reality of $\gamma=(1-v^2)^{-1/2}$ requires
$0\leq v\leq 1$.

Integrating twice (\ref{rot:EqMotion}) we get the function $H(v)$
that describes the profile of the spinning lumps,
\beq
 H(v)=\int_{v_o}^{v}\frac{-f(x)}{(g(x)^2-f(x)^2)^{\frac{1}{2}}}
dx\,;\qquad {\rm with} \quad f(v)=2k-3(v^2+2c)\gamma^{-3}, \qquad
g(v)=6\widetilde{\omega}_\phi v \gamma^{-3}.
 \label{Hfunction}
 \eeq
where $c$ is an integration constant. Plasma rings have also a inner
surface: its profile is described by a similar expression obtained
by multiplying the first relation in (\ref{Hfunction}) by $-1$ and
replacing $v_o \rightarrow v_{\rm i}$ (with $v_{\rm i}$ being the
inner velocity; for details see \cite{Bhardwaj:2008if}).

To make contact again with \cite{Bhardwaj:2008if} we define the
dimensionless energy, angular momentum and entropy as
\eq \widetilde{E} = \frac{\rho_0^2 E}{4\pi \sigma^3}, \qquad
\widetilde{J} = \frac{\rho_0^3 J}{20\pi  \sigma^4}, \qquad
\widetilde{S} = \frac{\rho_0^{\frac{11}{5}} S}{20\pi
\alpha^{\frac{1}{5}}\sigma^3}\,,\label{DlessCharge}
 \eeq
where $E,J,S$ follow from (\ref{Var:charges}) and
(\ref{Var:Entropy}). It is also useful to introduce the following
functions:
\begin{eqnarray}
\label{tubeeng}
 \widetilde{E}_t (v_o) &=&\frac{\widetilde{L}}{12\widetilde{\omega}_\phi^2}
\left(2k\gamma_o^5\lp
v_o^2+2-2\gamma_o^{-5}\rp+3v_o(v_o+\widetilde{\omega}_\phi)\right)  \nonumber\\
 \widetilde{J}_t (v_o)&=&\frac{\widetilde{L}k\gamma_o^5}{30\widetilde{\omega}_\phi^3}
\left(2\gamma_o^{-5}-2+5v_o^2\right)\,,\qquad \widetilde{S}_t (v_o)
=\frac{\widetilde{L}k^{4/5}}{6\widetilde{\omega}_\phi^2}
(\gamma_o^3-1).
\end{eqnarray}
and
\begin{eqnarray}
\label{balleng}
 \widetilde{E}_b(a,b) &=& \frac{1}{\widetilde{\omega}_\phi^3}
\int_{a}^{b}dv\lp v H(v)\lp k(4+v^2)\gamma^7 +1\rp+
    \widetilde{\omega}_\phi v \sqrt{1+H'(v)^2}\rp,
    \nonumber\\
  \widetilde{J}_b(a,b) &=& \frac{k}{\widetilde{\omega}_\phi^4}
    \int_{a}^{b}v^3\gamma^7 H(v) dv\,,\qquad
  \widetilde{S}_b(a,b) =
  \frac{k^{4/5}}{\widetilde{\omega}_\phi^3} \int_{a}^{b} v \gamma^5
  H(v)dv\,.
\end{eqnarray}

We now discuss the seven families of axisymmetric spinning plasma
lumps that we found to be solutions of (\ref{rot:EqMotion}) (we
present the ones with non-trivial profiles in Figs.~\ref{fig:PT},
\ref{fig:PB}, and \ref{fig:PR}. The figures should be rotated along
the vertical axis $v=0$):
\begin{figure}[th]
\centerline{\includegraphics[width=.193\textwidth]{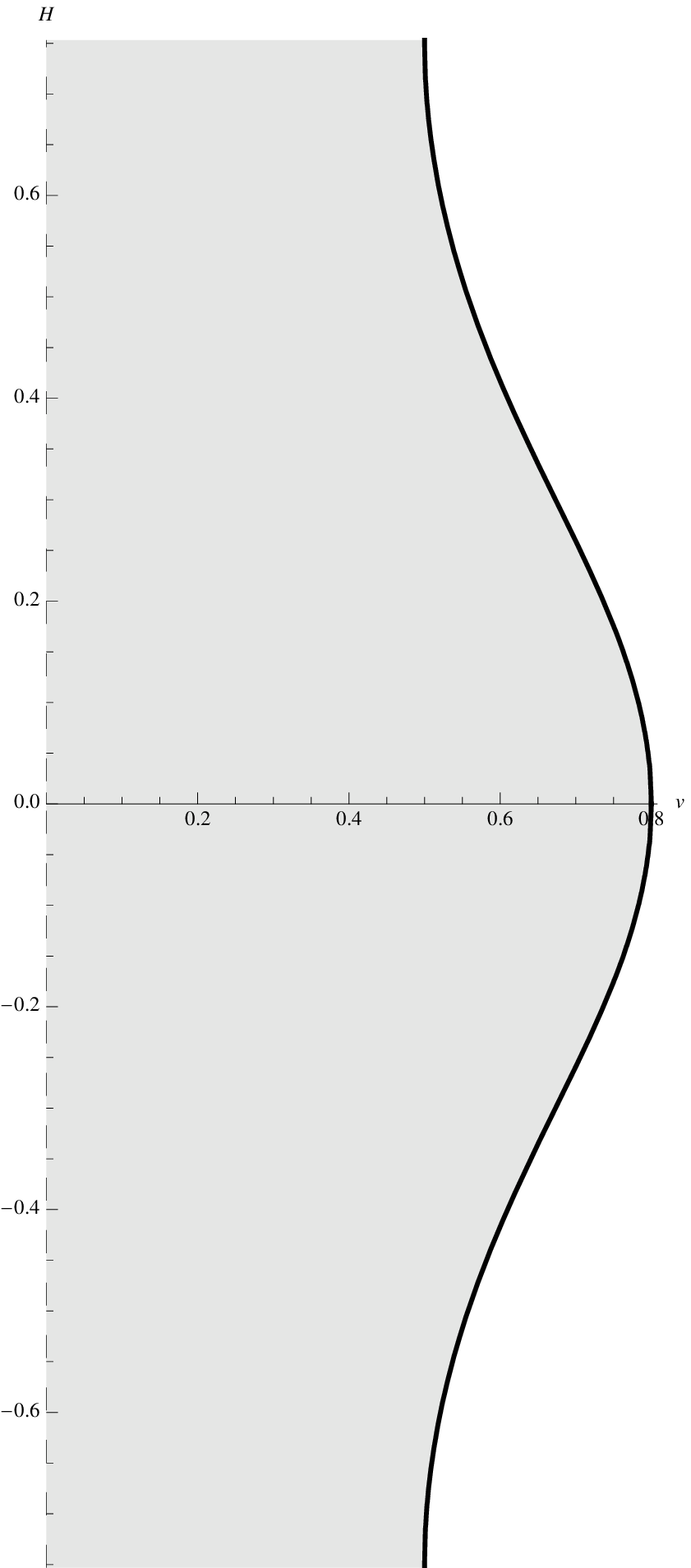}
\hfill
\includegraphics[width=.22\textwidth]{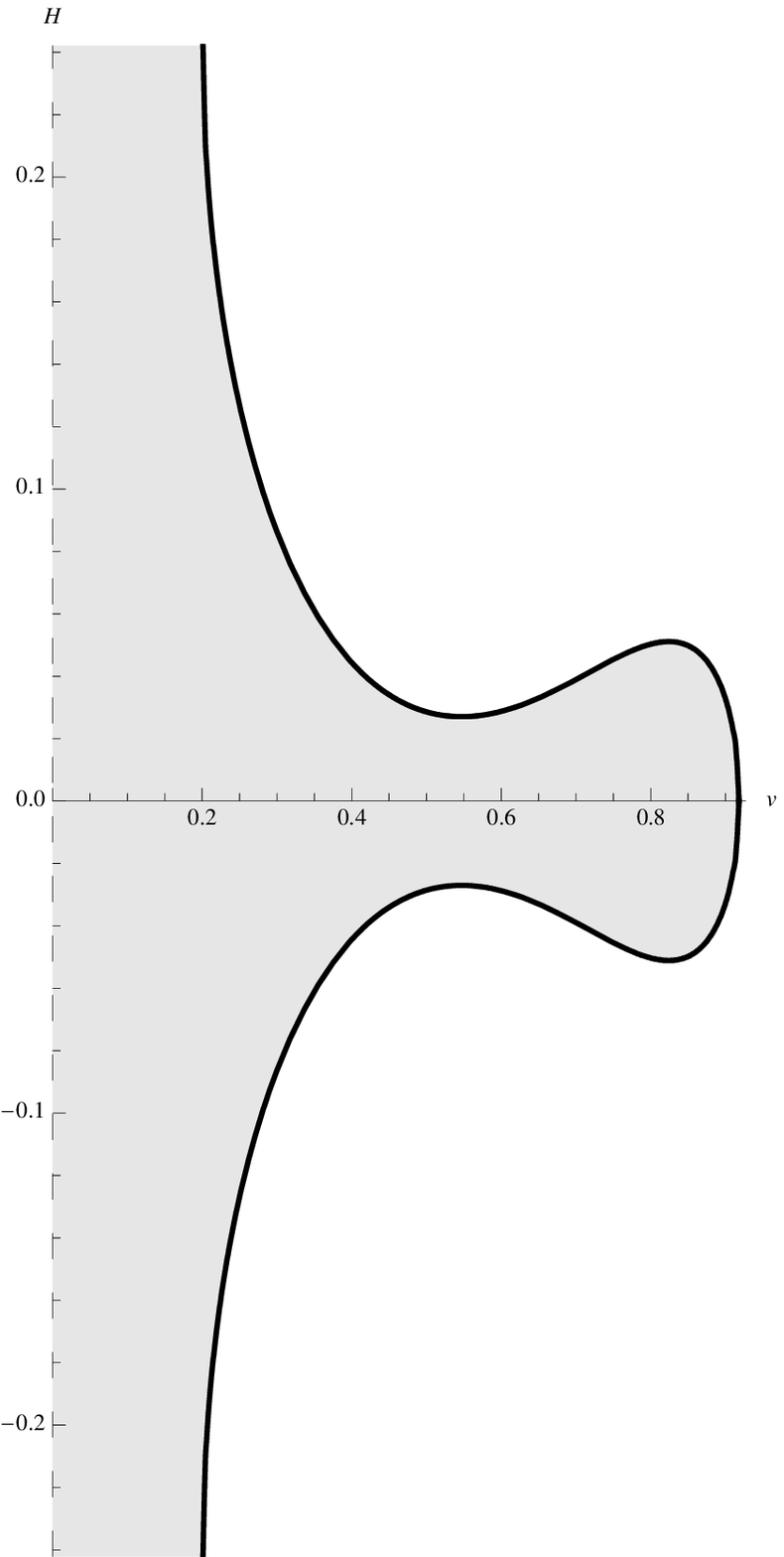}
\hfill\includegraphics[width=.22\textwidth]{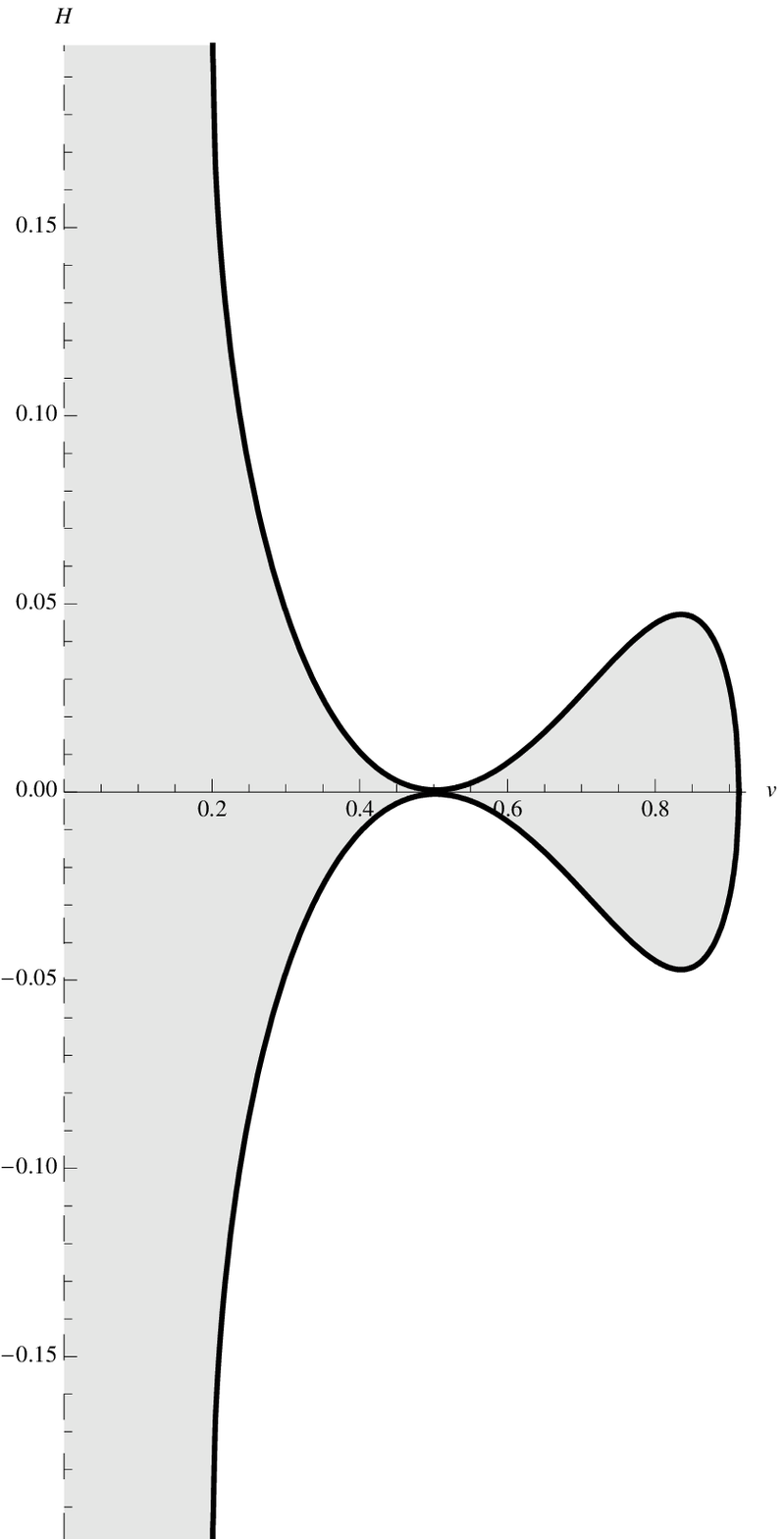}}
\caption{\small Figures of equilibrium of a rotating plasma tube.
From left to right, an ordinary non-uniform rotating plasma tube ($v_m=.5$,
$v_o=.8$, $k=.628024$), a pinched plasma tube ($v_m=.201016$,
$v_o=.918$, $k=.15$) and a plasma tube on the verge of splitting off
a plasma ring ($v_m=.201016$, $v_o=.9133$, $k=.15$). The parameters
have been chosen such that the three tubes share the same
periodicity.} \label{fig:PT}
\end{figure}
\begin{figure}[t]
\centerline{\includegraphics[width=.22\textwidth]{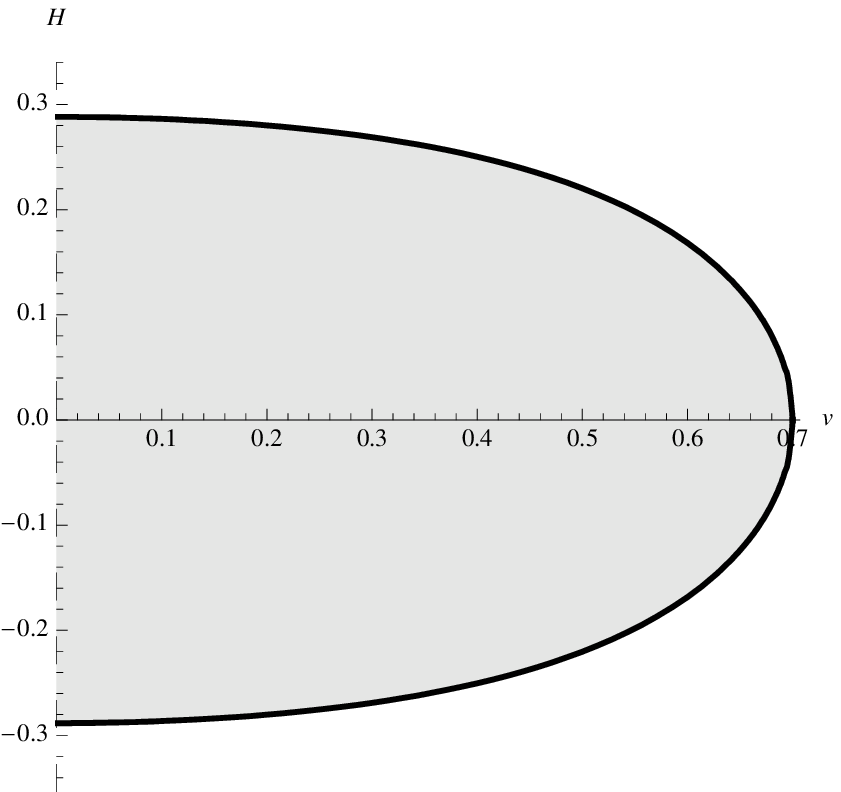}
\hfill\includegraphics[width=.22\textwidth]{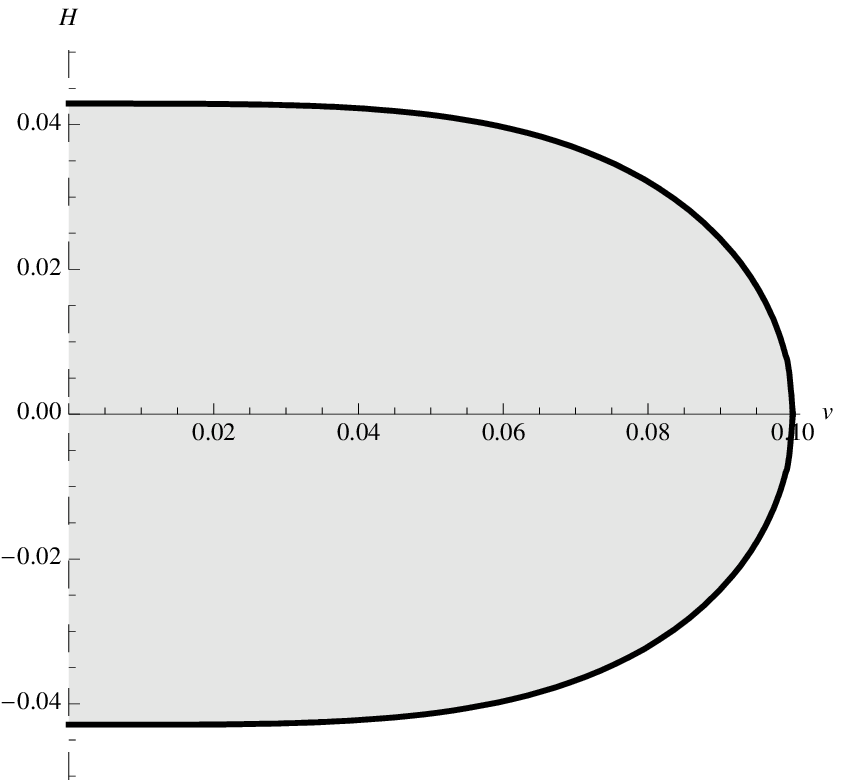}\hfill\includegraphics[width=.22\textwidth]{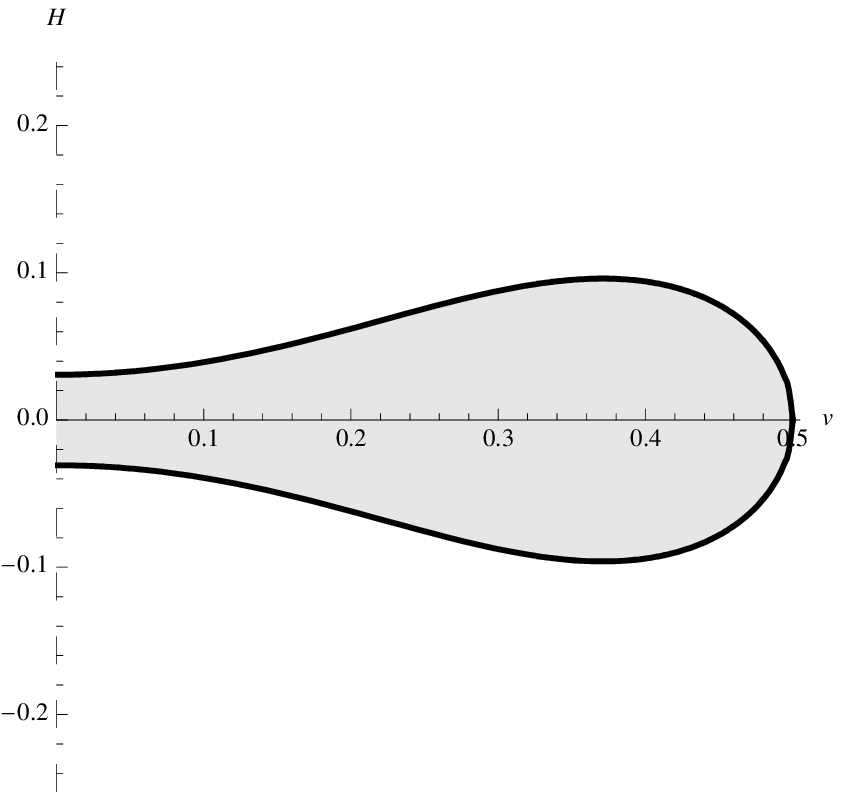}
\hfill\includegraphics[width=.22\textwidth]{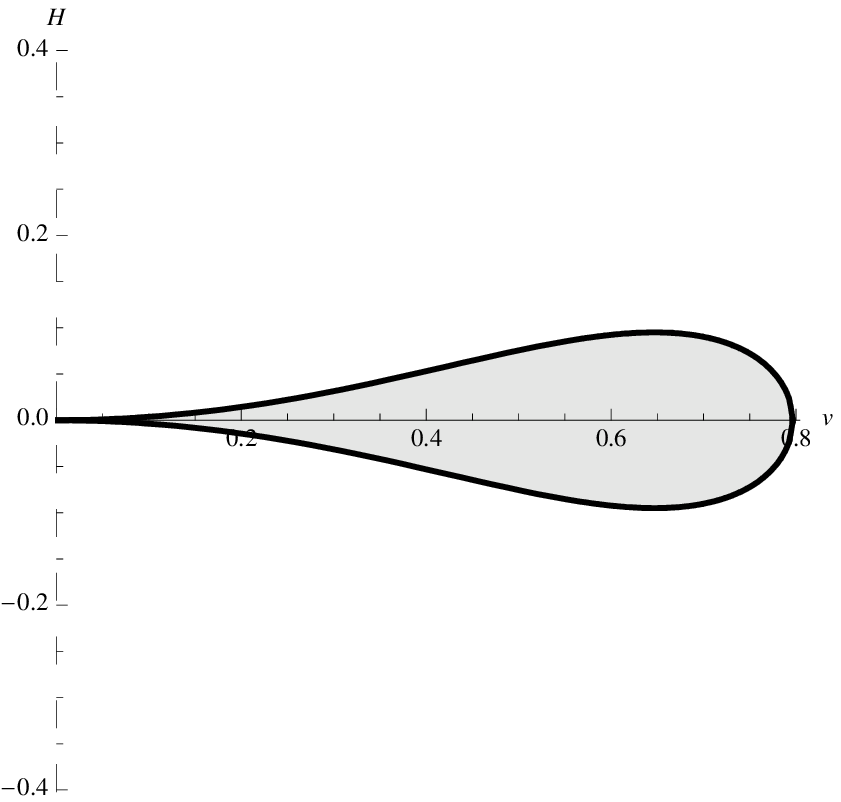}}
\caption{\small Figures of equilibrium of a rotating plasma ball.
From left to right, an ordinary rotating drop ($v_o=.7$, $k=2$), a
marginal rotating drop ($v_o=.1$, $k=21$), a pinched plasma ball
($v_o=.5$, $k=.83$) and a pinched ball on the verge of splitting of
a plasma ring ($v_o=.7965$, $k=.5$).} \label{fig:PB}
\end{figure}
\begin{figure}[th]
\centerline{\hfill
\includegraphics[width=.38\textwidth]{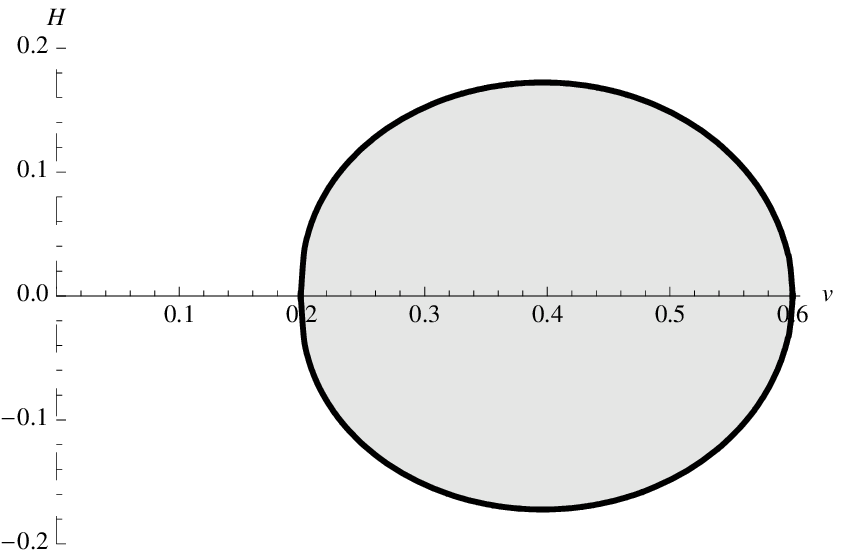}
\hfill\includegraphics[width=.4\textwidth]{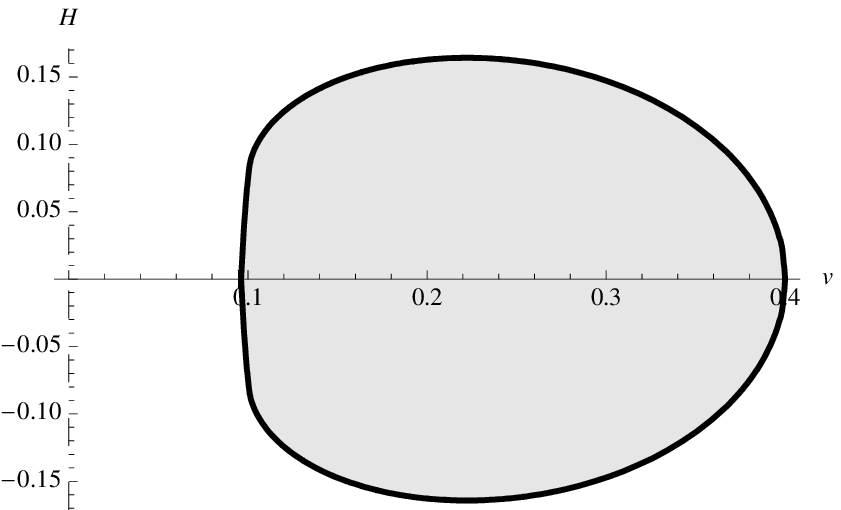}\hfill}
\caption{\small Figures of equilibrium of a rotating plasma ring. On
the left, a thin plasma ring ($v_{\rm i}=.2$, $v_o=.6$, $k=9$) and
on the right a fat plasma ring ($v_{\rm i}=.1$, $v_o=.4$, $k=2$).}
\label{fig:PR}
\end{figure}
\begin{itemize}
\item{\it Uniform plasma tubes (UT):}
These have constant radius $R_o$ along their length $L$. They have
$\widetilde{\omega}_\phi=v_o(k \gamma_o^5-1)$, where
$\gamma_o\equiv\gamma{\bigl |}_{v_o}$. Their energy, angular
momentum and entropy are
$\widetilde{E}=\widetilde{E}_t(v_o)$,
$\widetilde{J}=\widetilde{J}_t(v_o)$ and
$\widetilde{S}=\widetilde{S}_t(v_o)$ as
defined in (\ref{tubeeng}).

\item{\it Uniform hollow tubes (UHT):}
These have two constant radii, outer $R_o$ and inner $R_{\rm i}$,
along their length $L$ delimiting the tubular region inside which
the fluid lies. The Young-Laplace equation on the inner and outer
boundaries fixes a constraint between these two radii, which can be
written as $\widetilde{\omega}_\phi=v_o(k \gamma_o^5-1)=
 v_{\rm i}(1-k \gamma_{\rm i}^5)$,
 where $\gamma_{o,{\rm i}}\equiv\gamma{\bigl |}_{v_{o,{\rm i}}}$.
The energy, angular momentum and entropy are
$\widetilde{E}=\widetilde{E}_t(v_o)-\widetilde{E}_t(v_{\rm i})$,
$\widetilde{J}=\widetilde{J}_t(v_o)-\widetilde{J}_t(v_{\rm i})$ and
$\widetilde{S}=\widetilde{S}_t(v_o)-\widetilde{S}_t(v_{\rm i})$,
with the definitions in (\ref{tubeeng}).

\item{\it Non-uniform plasma tubes (NUT):} These tubes have a radius
that varies from a non-vanishing minimum, $v_m$, up to a maximum,
$v_o$, as we go along their length. At both these extreme points we have
$H'(v)=-\infty$, and in between them $H'(v)$ is always negative (so the
point in between the two extreme points where $H''(v)$ changes
sign has $H'(v)<0$); see Fig.~\ref{fig:PT}.a. In the critical NUT
configuration, $H'(v)=0$ at the inflection point and the NUT and
pinched NUT (discussed next) meet in what we call the critical NUT.
These boundary conditions imply the following relations between the
parameters,
\begin{eqnarray}
\label{NUTparameters}
 c&=&\frac{k}{3}\lp 1-v_m^2 \rp^{-3/2}
-\frac{1}{2}v_m^2 - \widetilde{\omega}_\phi v_m \,,\nonumber\\
\widetilde{\omega}_\phi &=&
\frac{1}{6v_o}\lp1-\frac{v_m}{v_o}\rp^{-1}
 \lpp 2k\lp \lp 1-v_o^2\rp^{-3/2}-\lp 1-v_m^2\rp^{-3/2}\rp -3\lp v_o^2-v_m^2 \rp
 \rpp .
\end{eqnarray}
The energy, angular momentum and entropy of the NUTs are given,
respectively, by
$\widetilde{E}=\widetilde{E}_t(v_m)+\widetilde{E}_b(v_m,v_o)$,
$\widetilde{J}=\widetilde{J}_t(v_m)+\widetilde{J}_b(v_m,v_o)$ and
$\widetilde{S}=\widetilde{S}_t(v_m)+\widetilde{E}_b(v_m,v_o)$ where
we use the functions defined in (\ref{tubeeng}) and (\ref{balleng}).

\item{\it Pinched non-uniform plasma tubes (pNUT):}
These tubes also have a radius that varies from a non-vanishing
minimum, $v_m$, up to a maximum, $v_o$. At both these extreme
points we have $H'(v)=-\infty$. But, contrary to the NUT solutions, $H'(v)$
now changes sign twice as $v$ runs from $v_m$ to $v_o$ (so the
point in between the two extreme points where $H''(v)$ changes
sign has now $H'(v)>0$), see Fig.~\ref{fig:PT}.b. These boundary
conditions still imply the relations (\ref{NUTparameters}) and the
conserved charges of the pinched NUTs are given by the same
functional defined for the NUT in the previous item. As the rotation
increases we reach a configuration where we have a pNUT and a
plasma tube on the verge of splitting off a plasma ring; see
Fig.~\ref{fig:PT}.c.

\item{\it Plasma balls (B):} These plasma configurations are oblate spheres.
At the extreme points they satisfy $H'(v=0)=0$ and
$H'(v=v_o)=-\infty$. Moreover, $H'(v)$ is always negative at the
boundary as we go from $v=0$ up to $v=v_o$; see Fig.~\ref{fig:PB}.a.
The condition $H'(0)=0$ fixes $c=k/3$ and $H'(v_o)=-\infty$ requires
$f(v_o)=g(v_o)$ which fixes $\widetilde{\omega}_\phi=
\frac{1}{6v_o}\left(2k\gamma_o^3-2k-3v_o^2\right)$. Finally, the condition
$H'(v)<0$ in the interval $(0,v_o)$ implies $k>1$. For $k=1$ the critical
ball configuration is reached (see Fig.~\ref{fig:PB}.b). Here, the
plasma ball joins the pinched ball branch discussed next. The
conserved charges are given, respectively, by
$\widetilde{E}=\widetilde{E}_b(0,v_o)$,
$\widetilde{J}=\widetilde{J}_b(0,v_o)$ and
$\widetilde{S}=\widetilde{E}_b(0,v_o)$ where we use the functions
defined in (\ref{balleng}). The balls must have a distance between
their poles smaller than the compact dimension length. The same
applies to the solutions we discuss next.

\item{\it Pinched plasma balls (pB):}
These balls also have $H'(0)=0$ and $H'(v_o)=-\infty$ which again
requires $c=k/3$ and $\widetilde{\omega}_\phi=
\frac{1}{6v_o}\left(2k\gamma_o^3-2k-3v_o^2\right)$. But contrary to the plasma
balls, $H'(v)$ changes from positive to negative sign in the interval
$(0,v_o)$; see Fig.~\ref{fig:PB}.c. This requires $0<k<1$. However not
all $k$ in this interval give a physical solution. They must further
satisfy $H(v=0)>0$. At the critical situation where $H(0)=0$ the pinched
plasma ball branch joins the fat plasma ring phase discussed next (see
Fig.~\ref{fig:PB}.d). The conserved charges of the pinched balls are
given by the same expressions as for the
balls.

\item{\it Plasma rings (R):}
These annulus configurations are characterized by
$H^{\prime}(v_{\rm_i})=+\infty$ and $H^{\prime}(v_o)=-\infty$ at the
inner and outer boundary, respectively, and for having
$H(v_{\rm_i})=H(v_o)=0$. This requires that
 $f(v_{\rm_i})=-g(v_{\rm_i})$ and $f(v_o)=g(v_o)$. For fixed outer
boundary velocity, $v_o=\omega_\phi R_o$, there are two possible
inner velocities $v_{\rm_i}$ satisfying these conditions. The
smaller one describes the fat plasma ring (that joins the pinched
ball branch at the critical configuration where $v_{\rm_i}\rightarrow
0$) and the larger $v_{\rm_i}$ describes the thin ring (see
Fig.~\ref{fig:PR}). The two rings meet at a regular critical ring
when the two aforementioned $v_{\rm_i}$'s are equal. The energy,
angular momentum and entropy of the rings are given, respectively,
by $\widetilde{E}=\widetilde{E}_b(v_{\rm_i},v_o)$,
$\widetilde{J}=\widetilde{J}_b(v_{\rm_i},v_o)$ and
$\widetilde{S}=\widetilde{E}_b(v_{\rm_i},v_o)$ with
functions defined in (\ref{balleng}).
\end{itemize}

There also appears the possibility of {\it non-uniform hollow tubes
(NUHT)}, \ie
hollow tubes with an inhomogeneous profile along $z$, but we have not
obtained direct evidence for them. If they exist, it would be
interesting to increase their rotation and investigate their possible
pinches and connections to other phases.

Having identified these families of plasma solutions, it should be
interesting to represent their properties in a phase diagram where we
fix the periodic direction length and the energy of the solution, and
let the entropy vary with the angular momentum (alternatively, we could
also fix $\widetilde{L}$ and $\widetilde{J}$ and represent the solutions
in the $\widetilde{S}(\widetilde{E})$ diagram). This requires
substantial and delicate numerical work since this is a shooting problem
on three parameters. The curves for uniform tubes and balls with a fixed
conserved charge can be obtained relatively easily but the numerical
search of the non-uniform tubes is considerably more difficult. We leave
this for future work.

\setcounter{equation}{0}
\section{Rayleigh-Plateau instability of plasma tubes\label{sec:RP}}

Non-relativistic fluid jets and liquid bridges in between two disk
plates are unstable against the so-called Rayleigh-Plateau
instability if their length is larger than their transverse
perimeter. This instability is active even in the absence of gravity
and surface tension is the crucial mechanism responsible for it: in
the simplest case, capillarity forces conspire to reduce the surface
area of the fluid. This is a long wavelength instability that makes
the fluid cylinder pinch-off.

In this section we show that uniform plasma tubes are naturally
afflicted by the relativistic Rayleigh-Plateau instability and we
will study in detail its properties. We start by getting a good
understanding of the mechanism responsible for the instability
noting that there are perturbation modes that decrease the plasma
tube potential energy while keeping its volume fixed (so this
statement is independent of whether or not the fluid is
compressible, \ie of the equation of state). Equivalently, in view
of the results of section \ref{sec:Variation}, these modes are
unstable because they increase the entropy of the system while
keeping its conserved charges fixed. In subsection \ref{sec:thresh},
this approach will allow to find the minimum length below which the
uniform tube can be stable. Then in subsection \ref{sec:Dispersion}
we will find the dispersion relation for this instability.

In the dual gravitational system, uniform black strings are expected to
suffer from the gravitational Gregory-Laflamme instability. It is also a
long wavelength instability that makes black strings longer than the
horizon radius pinch. We will see that the plasma surface tension and
gravitational instabilities have similar properties.

\subsection{Unstable threshold mode from entropy or potential energy variation \label{sec:thresh}}

\subsubsection{Threshold mode for static and boosted plasmas}

Consider an axisymmetric perturbation on a uniform plasma tube,
\eq
 r(z)=R_o+\epsilon R_1\cos(kz)+\epsilon^2 R_2\,,\qquad \epsilon\ll 1\,. \label{st:pert}
 \eeq
We take the tube length to be given by the Rayleigh-Plateau
wavelength because that is where it might break, $L=\lambda=2\pi/k$.
The volume of the perturbed tube, up to a second order expansion, is
then
\eq
 V= \lambda \frac{\pi^{(n+1)/2}}{\Gamma\lp\frac{n+3}{2}\rp}\, R_o^{n-1}\left(R_o^2+\frac{n+1}{2}\lp \frac{n}{2}R_1^2+2R_oR_2
  \rp  \epsilon^2\right) +\mathcal{O}(\epsilon^3)\,.\label{stat:vol}
\eeq
We require that the volume of the tube is kept fixed up to second
order in the perturbation (equivalently, its conserved charges must
be kept fixed) so this constrains $R_2$ as a function of the lower
order radius,
\eq
 R_2=-\frac{n}{4}\frac{R_1^2}{R_o}\,. \label{stat:Rconst}
 \eeq

 The difference between the perturbed and
unperturbed surface tension potential energy is
\eq
 \Delta U_{\sigma}
 =\sigma \frac{2\pi^{(n+1)/2}}{4\Gamma\lp\frac{n+1}{2}\rp}\lp k^2R_o^2-n\rp R_1^2\epsilon^2+\mathcal{O}(\epsilon^3)
 \label{stat:PotVar}
\eeq
where we used (\ref{stat:Rconst}). Unstable modes are those that
decrease the potential energy, \ie surface area, of the fluid tube.
So the condition $\Delta U_{\sigma}\leqslant 0$ for fixed volume
requires
\eq k R_o\leqslant \sqrt{n}\,,\label{stat:threshold} \eeq
For static plasmas the threshold unstable mode is then $k =
\sqrt{n}/R_o$ and it increases with the spatial dimension. Finally,
a boost along the $z$-direction has only a kinematical effect on
this instability, Lorentz-contracting the threshold wavelength
\cite{Hovdebo:2006jy}.

\subsubsection{Threshold mode for rotating plasma and non-axisymmetric modes\label{ss:thresholdrotation}}

We  now generalize the previous study to rotating and boosted
uniform plasma tubes, as well as non-axisymmetric perturbations on
it. Again, we will restrict to the $n=1$ dimensional case, since for
higher $n$, the profile of the drop has a polar angle dependence
that prevents an analytical study.

The general perturbation on the uniform plasma tube reads,
\eq
 r(z,\phi)=R_o+\epsilon R_1\cos(kz)\cos(m\phi)+\epsilon^2 R_2\,,\qquad \epsilon\ll 1\,. \label{rot:pert}
 \eeq
This perturbation includes axisymmetric modes ($m=0$) as well as
non-axisymmetric ones (integer $m\neq 0$). Again, we take the tube
length to be equal to the instability wavelength, $L=\lambda=2\pi/k$.

The volume of the perturbed tube, up to a second order expansion, is
\eq
 V= \lambda \pi R_o^2 +\frac{\lambda}{4}\left(\pi\lp R_1^2+8R_oR_2
  \rp  + R_1^2\,\frac{\sin(4m\pi)}{4m} \right)\epsilon^2 +\mathcal{O}(\epsilon^3)\,.\label{rot:vol}
 \eeq
We require that the volume of the tube is kept fixed up to second
order in the perturbation (equivalently, its conserved charges must
be kept fixed) so this constrains $R_2$ as a function of the lower
order radius,
\eq
 R_2=-\frac{R_1^2}{4R_o}\,,\quad{\rm for}\:\:m=0\,; \qquad
 R_2=-\frac{R_1^2}{8R_o}\,,\quad{\rm for}\:\:m\neq 0\,. \label{rot:Rconst}
 \eeq

The potential energy of the tube is the sum of the surface tension
and centrifugal contributions as defined in section
\ref{sec:Variation}. Using the constraint (\ref{rot:Rconst}), the
total potential energy difference between the perturbed and
unperturbed configurations $\Delta U=\Delta U_{\sigma}+\Delta U_{\rm
cf}$ is then given by ($\delta_m=1$ if $m=0$ and vanishes otherwise)
\eq
 \Delta U=\frac{\pi\lambda\sigma}{4R_o}
 \lp1+\delta_m\rp\left(\lp k^2R_o^2-1+m^2\rp
 -\frac{5}{4}\frac{\rho_*\omega_{\phi}^2R_o^3}{\sigma}\left(1-R_o^2\omega_{\phi}^2-\omega_z^2
  \right)^{-7/2}\right)R_1^2\epsilon^2+\mathcal{O}(\epsilon^3)\,.\label{rot:totPotVar}
 \eeq
The Rayleigh-Plateau instability is active for wavenumbers that decrease
the potential energy of the fluid tube. So the condition $\Delta
U\leqslant 0$ for fixed volume requires
\eq k^2R_o^2+m^2\leqslant 1
 +\frac{5}{4}\frac{\rho_*\omega_{\phi}^2R_o^3}{\sigma}\left(1-R_o^2\omega_{\phi}^2-\omega_z^2
  \right)^{-7/2}\,,\label{rot:threshold}
 \eeq
The inverse of the ratio ${\rho_*\omega_{\phi}^2R_o^3}/{\sigma}$,
that measures the competition between the surface tension and
centrifugal effects, is often called the rotating Bond number or
Hocking parameter. Rotation increases the critical wavenumber making
the plasma tube unstable for a wider range of wavelengths. Only
axisymmetric modes are unstable in the static case but when rotation
is added, non-axisymmetric modes can become also unstable for
sufficiently high velocity. We will discuss the dual gravitational
interpretation of these unstable modes in section~\ref{sec:RPGL}.

\subsection{Rayleigh-Plateau dispersion relation \label{sec:Dispersion}}

In this subsection we want to address the stability of a uniform
plasma tube when we perturb it. The dynamics of the perturbations
is dictated by the hydrodynamic equations, subject to appropriate
boundary conditions.

Perturbations take the plasma away from thermal equilibrium and therefore
viscosity and diffusion effects start to contribute. The energy-momentum
tensor of the fluid includes now not only the
perfect fluid and the boundary surface tension terms
(\ref{GenLumpTuv}), but also a dissipative contribution. As we will
show, the uniform plasma tube is afflicted by the Rayleigh-Plateau
instability and surface tension is the mechanism responsible for it.
Viscosity and diffusion play no role on the activation
of the instability and have a subleading effect on the dispersion
relation, simply correcting the threshold wavelength and the time-scale of the instability. Hence, in our analysis we shall neglect the dissipation
contribution to the fluid energy-momentum tensor, and simply comment at the
end the subleading effects it introduces.

To study the stability of rigidly rotating uniform plasma tubes, we
consider a generic unperturbed uniform tube with velocity
(\ref{velocity}). The precise expression for the unperturbed plasma
pressure and density is fixed once we specify the equation of state.
In particular, for the plasma we are interested in, these are
specified by equations (\ref{ConfEqState}) and (\ref{DpEquil}). The
particular choice of the equation of state will however not be
fundamental in our analysis.

A generic perturbation on the uniform tube is described as
\eq
 u^\mu=u^\mu_{(0)}+\delta u^\mu\,,\qquad P=P_{(0)}+\delta P\,,
 \qquad \rho=\rho_{(0)}+\delta \rho\,,\label{perturbation}
 \eeq
where we denote an unperturbed quantity by $Q_{(0)}$, and the
perturbation as $\delta Q$. The perturbed state (\ref{perturbation})
must satisfy the relativistic continuity and Navier-Stokes equations,
(\ref{continuity}) and (\ref{NavierS}). The Young-Laplace equation
(\ref{YoungLap}) provides then a boundary condition for the perturbed
problem. After eliminating the $0^{\rm th}$ order terms using the
unperturbed hydrodynamic equations, the continuity and the Navier-Stokes
equations yield, up to first order in the perturbation,
\begin{eqnarray}
&&u^\mu_{(0)}\partial_{\mu}\delta\rho+\delta
u^{\mu}\partial_{\mu}\rho_{(0)} + (\rho_{(0)}+P_{(0)}) \nabla_{\mu}
\delta u^\mu + (\delta\rho+\delta P) \nabla_{\mu} u^\mu_{(0)} = 0\,,
\label{Pert:continuity}\\
&&(\rho_{(0)}+P_{(0)}) \lp \delta u^\mu \nabla_{\mu}
u^\nu_{(0)} + u^\mu_{(0)} \nabla_{\mu} \delta u^\nu \rp +
(\delta\rho+\delta P) u^\mu_{(0)}
\nabla_{\mu} u^\nu_{(0)} \nonumber\\
&&\qquad\quad+ \lp g^{\mu\nu}+u^\mu_{(0)} u^\nu_{(0)} \rp \nabla_{\mu} \delta
P+ \lp  \delta u^\nu u^\mu_{(0)} + u^\nu_{(0)} \delta u^\mu\rp
\nabla_{\mu} P_{(0)}=0 \,. \label{Pert:NavierS}
\end{eqnarray}
The density and pressure perturbations are not independent. They are
related by the equation of state (\ref{ConfEqState}), valid also out
of equilibrium. In particular, perturbation of the first relation in
(\ref{ConfEqState}) yields
\eq \delta \rho= (n+3)\delta P \,. \label{rhoPpert}
 \eeq

Since we can expand any perturbation in a Fourier series, we
restrict the analysis to a generic mode and consider a perturbation
that disturbs the boundary surface according to
\eq r=R(t,z,\phi)\,, \qquad R(t,z,\phi)=R_o+ \epsilon \,e^{\omega
t}e^{ikz+im \phi} \,,\qquad \epsilon \ll R_o \,, \label{Pert:radius}
\eeq
where $R_o$ is the unperturbed radius of the uniform tube (note that
for the rotating $n>1$ case it is $\theta$-dependent). Positive
$\omega$ describes an instability with wavenumber $k$. These modes
break axisymmetry if $m\neq0$. The unit normal of
(\ref{Pert:radius}) is
\eq n_{\mu}=|n|^{-1}\lp -R_t'\delta_{\mu}^{\:t}+
\delta_{\mu}^{\:r}-R_{\phi}'\delta_{\mu}^{\:\phi}-R_z'\delta_{\mu}^{\:z}
\rp \,,\qquad |n|=\lp
1-R_t^{\prime\,2}+\frac{R_\phi^{\prime\,2}}{r^2}+R_z^{\prime\,2}
\rp^{1/2} . \label{Pert:unitnormal} \eeq
Naturally, we look for perturbations of the fluid quantities that
have the same form as the boundary disturbance,
\eq
\delta Q(t,r,z,\phi)=\delta Q(r) e^{\omega t}e^{ikz+im \phi}
\,,\qquad \delta Q\equiv\{\delta u^\mu,\delta P,\delta \rho \} \,.
\label{Pert:Q}
\eeq

The perturbed hydrodynamic equations (\ref{Pert:continuity}) and
(\ref{Pert:NavierS}) must be supplemented by appropriate boundary
conditions. The first one demands normal stress balance on the
boundary. This means that the pressure perturbation that solves
(\ref{Pert:continuity}) and (\ref{Pert:NavierS}) must also satisfy
the third perturbed hydrodynamic equation, namely the equation that
follows from perturbing the Young-Laplace equation (\ref{YoungLap}),
\eq {\rm BC}\:\: {\rm I:}\qquad \delta P{\bigl |}_{\rm bdry}\simeq
\sigma \left[ K{\bigl |}_{R(t,z,\phi)} -K{\bigl |}_{R_o} \right] -
\left[\lp P_<^{(0)}-P_>^{(0)}\rp {\bigl |}_{R(t,z,\phi)} -  \lp
P_<^{(0)}-P_>^{(0)}\rp {\bigl |}_{R_o}\right] \,, \label{Pert:bc2}
\eeq
where on the rhs we evaluate the expression at the perturbed
boundary $r=R(t,z,\phi)$ defined in (\ref{Pert:radius}) and subtract
the unperturbed contribution evaluated at $r=R_o$. For our plasma,
$P_<^{(0)}-P_>^{(0)}$ is obtained from (\ref{DpEquil}), and the
extrinsic curvature is given from its definition (\ref{YoungLap})
using the unit normal (\ref{Pert:unitnormal}).

The second boundary condition is a kinematic condition requiring
that the normal component of the fluid velocity on the boundary
satisfies the perturbed version of (\ref{constraintYL}),
 $u^{\mu}_{(0)}\,\delta n_\mu +\delta u^\mu n_{\mu}^{(0)}=0$, where
 $\delta n_\mu\equiv n_\mu{\bigl |}_{R(t,z,\phi)}-n_{\mu}^{(0)}$ and
the unperturbed normal is $n_\mu^{(0)}\equiv n_\mu{\bigl
|}_{R_o}=\delta_\mu^r$. This ensures that the velocity perturbation
leaves the fluid confined inside the boundary. This boundary
condition then reads
\eq {\rm BC}\:\: {\rm II:}\qquad \delta u^{r}{\bigl |}_{\rm
bdry}\simeq \lp 1-R_o^2\omega_\phi^2-\omega_z^2\rp^{-\frac{1}{2}}
 \lp \omega+im\omega_\phi+ik\omega_z\rp \,\epsilon \,e^{\omega t}e^{ikz+im \phi}\,. \label{Pert:bc1}
\eeq

We have now all the ingredients needed to find perturbations of the form
(\ref{Pert:radius}) and (\ref{Pert:Q}) that might develop an
instability on the tube. We will find that there is indeed a long
wavelength instability known in non-relativistic systems as the
Rayleigh-Plateau instability (see for example \cite{Chandrasekhar:1981}).
It afflicts our plasma tube when, roughly, its length is larger than its transverse radius. We will first analyze this instability and its dispersion
relation $\omega(k)$ for a static uniform tube in any dimension $n$.
This is the simplest case where the crucial ingredients necessary to
activate the instability are present. Then we will study the changes introduced
in the instability properties by rotation and boost, as well as the effects of
non-axisymmetric modes. This will be done in the $n=1$ case, where the analysis can be done analytically, but we expect the results to hold qualitatively also for $n>1$. Finally, we will discuss briefly the subleading effects that viscosity
would introduce in the dispersion relation.

\subsubsection{Perturbations of static tubes in any dimension}

\begin{figure}[t]
\centerline{\includegraphics[width=.48\textwidth]{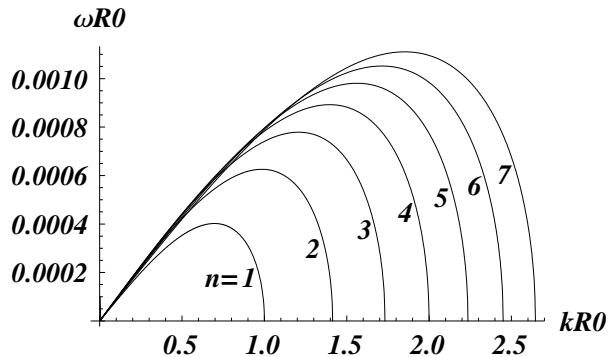}}
\caption{\small Plot of the dimensionless dispersion relation
$\omega(k)$ for the Rayleigh-Plateau instability in a static uniform
tube for several spacetime dimensions $d=n+3$. The instability
strength and threshold wavenumber increase as the spacetime
dimension grows. Intuitively, this is because the ``same"
perturbation decreases more the surface area of the plasma boundary
(increases more the plasma entropy) as the dimension increases. Note
also that the most unstable mode satisfies $\omega R_o{\bigl |}_{\rm
max} \gg \frac{\sigma}{\rho_0 R_o}$, and is thus within the regime
of validity of the hydrodynamic description (see section
\ref{sec:RegimeValidity}). This plot is to be qualitatively compared
with Fig.~3 of \cite{Gregory:1994bj} for the GL instability.}
\label{fig:GLstatic}
\end{figure}

Consider a static uniform plasma tube with unperturbed radius $R_o$
and $u^{\mu}_{(0)}=\delta^{\mu}{}_{t}$, in $(n+3)$ flat spacetime
dimensions, and deform its boundary by axisymmetric perturbations
($m=0$) of the form (\ref{Pert:radius}). The perturbed continuity
and Navier-Stokes equations, (\ref{Pert:continuity}) and
(\ref{Pert:NavierS}) have unstable modes that correspond to the
Rayleigh-Plateau instability, as we show in the sequel. Solving
(\ref{Pert:NavierS}) with perturbations (\ref{Pert:Q}), we find that
the non-vanishing $\delta u^\mu$ are (at higher order other
components will arise)
\eq \delta u^r (r)=- \omega^{-1} \lp \rho_{(0)}+P_{(0)}\rp^{-1}
\frac{d\, \delta P(r)}{dr} \,,\qquad \delta u^z
(r)=-i\,\frac{k}{\omega}\lp\rho_{(0)}+P_{(0)}\rp^{-1}\delta P(r)
 \,,
\label{Pert:Q:static}
\eeq
which, replaced in the continuity equation (\ref{Pert:continuity}), give
\eq \frac{d^2 \delta P(r)}{dr^2}+\frac{n}{r}\frac{d\, \delta
P(r)}{dr}-p^2 \delta P(r)=0\,, \qquad p=k\lp
1+(n+3)\frac{\omega^2}{k^2}\rp^{\frac{1}{2}}\,. \label{Pert:Bes:st}
\eeq
The $\omega$ term in the definition of $p$ arises from the first
term in (\ref{Pert:continuity}) proportional to $\delta\rho$ and use
of the equation of state (\ref{rhoPpert}). Equation
(\ref{Pert:Bes:st}) is a modified Bessel equation, whose solutions
are the modified Bessel functions of the first kind $I_{\pm
\frac{n-1}{2}}(p r)$ and second kind $K_{\frac{n-1}{2}}(p r)$. Of
these, $I_{-\frac{n-1}{2}}(p r)$ and $K_{\frac{n-1}{2}}(p r)$
diverge as $r^{-\frac{n-1}{2}}$ as $r\rightarrow 0$ and we discard
them. Therefore, the regular solution of (\ref{Pert:Bes:st}) at the
origin is
\begin{eqnarray}
&& \delta P(r)=A r^{-\frac{n-1}{2}} I_{\frac{n-1}{2}}(p r)\,, \nonumber\\
&& \delta u^r (r)=-\frac{A}{\omega\lp \rho_{(0)}+P_{(0)}\rp}\,
 r^{-\frac{n-1}{2}} \lp p\, I'_{\frac{n-1}{2}}(p r) -\frac{n-1}{2r}\, I_{\frac{n-1}{2}}(p r) \rp
 \,,\nonumber\\
&& \delta u^z(r)=-\frac{i k A}{\omega\lp
\rho_{(0)}+P_{(0)}\rp}\,r^{-\frac{n-1}{2}} I_{\frac{n-1}{2}}(p r)\,,
 \label{Pert:BesSol:st}
\end{eqnarray}
where we defined  $I'_{\nu}(x)\equiv \partial_x I_{\nu}(x)$, and where
$A$ is a constant that is fixed by the boundary condition (\ref{Pert:bc2})
to the leading-order value
\eq A\simeq \epsilon\sigma
\frac{R_o^{\frac{n-1}{2}-2}}{I_{\frac{n-1}{2}}(p R_o)}\lp k^2
R_o^2+\omega^2 R_o^2-n \rp  \,.
 \label{Pert:Ae:st}
\eeq
With this knowledge, and using the relations $I'_{\nu}(x)=I_{\nu
+1}(x)+\frac{\nu}{x}I_{\nu}(x)$ and (\ref{DpEquil}), the boundary
condition (\ref{Pert:bc1}) yields the desired dispersion relation
$\om=\omega (k)$, which reads
\eq \omega^2=\frac{n+3}{n+4}\,\frac{\sigma}{\rho_* R_o^3} \,\frac{p
R_o\, I_{\frac{n+1}{2}}(p R_o)}{I_{\frac{n-1}{2}}(p R_o)} \lp n -
k^2 R_o^2- \omega^2 R_o^2 \rp\,, \qquad p=k\lp
1+(n+3)\frac{\omega^2}{k^2}\rp^{\frac{1}{2}}.
 \label{Pert:Dispersion:st}
\eeq
A plot of this dispersion relation $\omega(k)$ for several values of
the dimension $n$ is shown in Fig.~\ref{fig:GLstatic}. To obtain
these plots we have needed to specify the value of
$\frac{\sigma}{\rho_* R_o}$. This amounts to considering plasma tubes of
a specific width $R_o$ relative to the length scale $\sigma/\rho_*$, which is
essentially the mean free path of the plasma. In principle we should
vary this width and study the dispersion relation as a
function of it.
The shape of the dispersion relation is indeed modified as
$\frac{\sigma}{\rho_* R_o}$ grows to values of order one. However, the
validity of the fluid description requires that we consider small
$\frac{\sigma}{\rho_* R_o}$ (see section~\ref{sec:RegimeValidity}). It
turns out that in this case the qualitative aspects of the dispersion
relation hardly change for different values of $\frac{\sigma}{\rho_*
R_o}$ and the
shapes shown in Fig.~\ref{fig:GLstatic} are
generic.
For definiteness, we have
taken
\beq\label{approxsig}
\frac{\sigma}{\rho_* R_o}=\frac{12}{n+6} 10^{-6}\,,
\eeq
so as to produce the plot. But we insist that the dimension-dependence in this
choice, and hence in the curves in the plot, is completely arbitrary: it only amounts
to choosing tubes of a
particular width in each dimension.

The Rayleigh-Plateau instability is active when $\omega>0$. The
instability gets stronger as the spacetime dimension increases since
the threshold wavenumber, as well as the wavenumber and frequency
where the maximum instability occurs, increase when $n$ grows.

There
are two special points that we can read easily from
(\ref{Pert:Dispersion:st}). One is the point where the curves cross
$\omega=0$, which gives
the threshold wavenumber below which the Rayleigh-Plateau
instability is active: $kR_o=\sqrt{n}$. So unstable modes satisfy
condition (\ref{stat:threshold}), confirming the threshold mode that
we obtained previously from entropical or energetical arguments in
subsection~\ref{sec:thresh}. The second point is at $k=0$, and
one can expand the Bessel functions in \eqref{Pert:Dispersion:st} to
check that $\omega$ at $k=0$
always vanishes, independently of $\frac{\sigma}{\rho_* R_o}$.

%
\subsubsection{Rotating and boosted plasma lumps \label{sec:RProtate}}

We want to consider now how rotation and boost of the fluid modify
the dispersion relation, and study the effect of non-axisymmetric
modes $(m\neq 0)$. We shall restrict to the $n=1$ case where an
analytical approach is viable. Take a uniform plasma tube rigidly
rotating with velocity (\ref{velocity}) and with unperturbed
constant radius $R_o$. The unperturbed state satisfies, once the
equation of state for the fluid is introduced, equations
(\ref{ConfEqState}) and (\ref{DpEquil}) with $n=1$.

Disturb now the plasma boundary with perturbations of the form
(\ref{Pert:radius}). To get the velocity perturbations we have to
solve (\ref{Pert:NavierS}) assuming perturbations of the form
(\ref{Pert:Q}). One finds, after use of (\ref{rhoPpert}), (in
(\ref{Pert:Q:rot0})-(\ref{Pert:cont:rot}) we restore the velocity of
light factor $c$ to make the perturbative expansion that we do next
more clear)
\begin{eqnarray}
&&  \delta u^t(r)=  \frac{\omega_\phi r^2}{c^2}\,\delta u^\phi(r)\,, \nonumber\\
 &&
 \delta u^r (r)=-\frac{4}{5\rho_*}\,
\frac{\varpi\gamma^{-1}}{\varpi^2+4\omega_{\phi}^2\chi}\lpp
\partial_r\,\delta P(r) +\lp \frac{2i\,m}{r}\frac{\omega_{\phi}}{\varpi}-3\gamma^2 r\,\frac{\omega_\phi^2}{c^2} \rp \delta
P(r) \rpp
,\nonumber \\
&& \delta u^\phi(r) =\frac{4}{5\rho_*}\,
\frac{\gamma^{-1}}{\varpi^2+4\omega_{\phi}^2\chi}  {\biggl[} \frac{2
\omega_{\phi}\chi}{r}
\partial_r\,\delta P(r)- \frac{i\,m\varpi}{r^2}\delta P(r)
-\frac{\omega_\phi\gamma^2}{c^2} \lp \varpi^2+10\omega_{\phi}^2\chi
\rp\delta P(r) {\biggr]}
\,, \nonumber \\
&& \delta u^z(r)= -i \,\frac{4}{5\rho_* \gamma}
\frac{k}{\varpi}\,\delta P(r)- \frac{4\gamma}{5\rho_* c^2} {\biggl[}
\omega_z \delta P  \nonumber \\
&& \hspace{2cm}-\frac{2\omega_\phi^2\omega_z
r}{\varpi^2+4\omega_{\phi}^2\chi} \lp \partial_r\,\delta P(r)
+\frac{2i\,m\varpi}{r}\frac{\omega_\phi}{\varpi}\delta P(r)-
\frac{3\omega_\phi^2\gamma^2r}{c^2}\delta P(r)\rp{\biggr]}
 \,,
\label{Pert:Q:rot0}
\end{eqnarray}
where we defined
\eq \varpi\equiv \omega+im\omega_\phi+ik\omega_z \,, \qquad
\chi\equiv 1+ \frac{\omega_\phi^2\gamma^2r^2}{c^2}\,.
\label{Pert:varpi:rot} \eeq
Note that (\ref{Pert:Q:rot0}) satisfies the perturbed version of
$u^\mu u_\mu=-c^2$, namely $u_\mu^{(0)}\delta u^\mu=0$. Equations
(\ref{Pert:Q:rot0}) can now be replaced in the continuity equation,
\eq \frac{d\,\delta u^r}{dr}+\frac{1}{r}\,\delta u^r + im\,\delta
u^\phi+ ik\,\delta u^z + \omega \,\delta u^t+ 4\gamma^2 r\,
\frac{\omega_\phi^2}{c^2} \delta u^r +\frac{16\varpi\gamma}{ 5\rho_*
c^2} \delta P=0 \,. \label{Pert:cont:rot} \eeq

As will become clear, it is not possible to get an analytical
expression for the dispersion relation unless we work in the small
velocity regime, $\omega_{\phi}R_o,\omega_z\ll c$. Since the
Rayleigh-Plateau instability is already present in the static case
and we just want to find if adding velocity to the solution
increases or decreases the instability strength and threshold
unstable mode, we will solve the hydrodynamic equations in this
regime (we will further represent higher order terms of the velocity
generically by $\mathcal{O}(\omega_i)$). This restriction is also
justified by the fact that the regime of validity of the
hydrodynamic description and thus of our results is restricted to
small rotation rates, as we will argue in section
\ref{sec:RegimeValidity}. So, to leading order (\ref{Pert:Q:rot0})
reduces to
\begin{eqnarray}
&& \delta u^t(r)=  \mathcal{O}(\omega_i^2)\,, \nonumber\\
 && \delta u^r (r)=-\frac{4}{5\rho_*}\,
\frac{\varpi}{\varpi^2+4\omega_{\phi}^2}\lp
\partial_r\,\delta P(r) +\frac{\omega_{\phi}}{\varpi}\,\frac{2i\,m}{r}\delta P(r) \rp + \mathcal{O}(\omega_i^2)\,,\nonumber \\
&& \delta u^\phi(r) =\frac{4}{5\rho_*}\,
\frac{\varpi}{\varpi^2+4\omega_{\phi}^2} \,\frac{1}{r} \lp
\frac{2 \omega_{\phi}}{\varpi} \partial_r\,\delta P(r) -\frac{i\,m}{r}\delta P(r) \rp + \mathcal{O}(\omega_i^2)\,, \nonumber \\
&& \delta u^z(r)= -i \,\frac{4}{5\rho_*} \frac{k}{\varpi}\,\delta
P(r) + \mathcal{O}(\omega_i^2)
 \,,
\label{Pert:Q:rot}
\end{eqnarray}
The perturbed continuity equation yields then to leading order in
the velocity,\footnote{Notice that the last term in
(\ref{Pert:cont:rot}) has a contribution $\omega/c^2$ that we keep
because our expansion is only on the velocities but not on the
frequency. This contribution is responsible for the $\omega^2/k^2$
term in $\eta$.}
\eq \frac{d^2 \delta P(r)}{dr^2}+\frac{1}{r}\frac{d\, \delta
P(r)}{dr}-\left[\eta^2+\frac{m^2}{r^2}\right] \delta
P(r)+\mathcal{O}(\omega_i^2) =0\,,\qquad \eta\equiv k\lp
1+\frac{4\omega^2}{k^2}\rp^{\frac{1}{2}}
 \lp 1+\frac{4\omega_{\phi}^2}{\varpi^2}\rp^{1/2}. \label{Pert:Bes:rot}
\eeq
 This is again a modified Bessel equation with solutions
$I_{m}(\eta r)$ and $K_{m}(\eta r)$ (note that if we had kept higher
order terms we would not be able to solve analytically the
differential equation). We discard the $K_{m}(\eta r)$ solution that
diverges at the origin and thus obtain the regular solution
\begin{eqnarray}
&& \delta P(r)\simeq A  I_{m}(\eta r)\,, \nonumber\\
&& \delta u^r (r)=
-\frac{4A}{5\rho_*}\frac{\varpi\eta}{\varpi^2+4\omega_{\phi}^2}\,
 \lp  I'_m(\eta r) -\frac{2i\,m}{\eta r}\,\frac{\omega_\phi}{\varpi}\, I_m(\eta r) \rp
 + \mathcal{O}(\omega_i)\,.
 \label{Pert:BesSol:rot}
\end{eqnarray}
The final expression for $\delta u^\phi$, $\delta u^z$ (that we do
not need) can then be obtained from (\ref{Pert:Q:rot}).

\begin{figure}[t]
\centerline{\includegraphics[width=.48\textwidth]{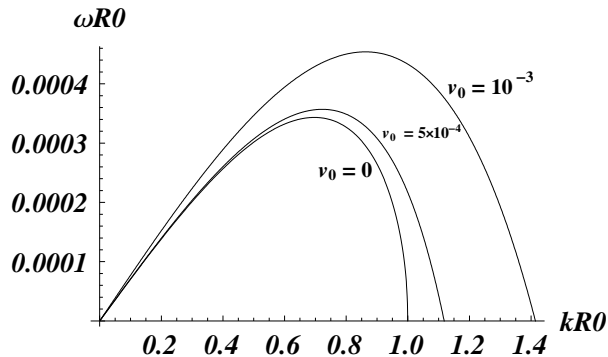}}
\caption{\small Plot of the dimensionless dispersion relation
$\omega(k)$ for the Rayleigh-Plateau instability in a rotating
uniform tube for dimensionless angular velocity $v_o=0; 5\times
10^{-4}; 10^{-3}$, where $v_o\equiv \omega_\phi R_o$. The
instability strength and threshold wavenumber increase as the
rotation grows. Intuitively, this is because the centrifugal force
scales with the radial distance and thus it is bigger in a crest
than in a trsough of the instability. The numerical data corresponds
to take $\frac{4\sigma}{5\rho_*  R_o}=10^{-6}$, $\omega_z=0$ and
$m=0$. The dispersion relation (\ref{Pert:Dispersion:rot}), valid in
the small velocity regime, is expect to be a good approximation up
to velocities of the order $v_o\sim 0.1-0.5$. Note also that the
most unstable mode satisfies $\omega R_o{\bigl |}_{\rm max} \gg
\frac{\sigma}{\rho_0 R_o}$, and is thus within the regime of
validity of the hydrodynamic description (see section
\ref{sec:RegimeValidity}). This plot should be compared with Fig.~1
of \cite{Kleihaus:2007dg} for the GL instability of rotating black
strings. } \label{fig:GLrotate}
\end{figure}

The boundary condition (\ref{Pert:bc2}) fixes the ratio $A/\epsilon$
to leading order as
\eq \frac{A}{\epsilon}\simeq \frac{\sigma R_o^{-2}}{I_m(\eta R_o)}
\left[ k^2 R_o^2+\omega^2
R_o^2-1+m^2-\frac{5}{4}\frac{\rho_*\omega_\phi^2R_o^3}{\sigma}\lp
1-\omega_\phi^2R_o^2 -\omega_z^2\rp^{-7/2}\right] \,.
 \label{Pert:Ae:rot}
\eeq
Use of this ratio on the boundary condition (\ref{Pert:bc1}) yields
finally the leading dispersion relation $\omega (k)$
\begin{eqnarray}
&&\hspace{-0.5cm} \varpi^2 \simeq -4\omega_{\phi}^2+
\frac{4\sigma}{5\rho_* R_o^3} \left[ \frac{\eta R_o\, I_{m+1}(\eta
R_o)}{I_m(\eta R_o)}+m \lp
1-i\,\frac{2\omega_\phi}{\varpi}\rp\right] \nonumber\\
&&\hspace{2.0cm}\times\lp
1-m^2+\frac{5}{4}\frac{\rho_*\omega_\phi^2R_o^3}{\sigma}\lp
1-\omega_\phi^2R_o^2 -\omega_z^2\rp^{-7/2}-\omega^2 R_o^2 -k^2 R_o^2
\rp \,.
 \label{Pert:Dispersion:rot}
\end{eqnarray}
A plot of this dispersion relation $\omega(k)$ for different values
of $\omega_\phi$ (and $\omega_z=m=0$) is shown in
Fig.~\ref{fig:GLrotate}. The dispersion relation
(\ref{Pert:Dispersion:rot}), valid in the small velocity regime, is
expect to be a good approximation up to velocities of the order
$v_o\sim 0.1-0.5$. The Rayleigh-Plateau instability is active when
$\omega>0$. The marginal unstable mode with $\omega=0$ satisfies
condition (\ref{rot:threshold}), confirming the threshold mode
obtained in subsection~\ref{ss:thresholdrotation}. As the rotation
$\omega_\phi$ is increased, the threshold wavenumber, as well as the
wavenumber and frequency where the maximum instability occurs, and
the instability grows stronger. Fixing $\omega_\phi \neq 0$,
increasing $\omega_z$ also makes the instability stronger. Finally
note that contrary to the static case where only axisymmetric modes
were unstable, now non-axisymmetric modes can also become unstable
for sufficiently high rotation.

Note that we decided to keep the higher order terms in velocities in
the boundary condition (\ref{Pert:Ae:rot}) and in
(\ref{Pert:Dispersion:rot}) instead of doing a Taylor expansion. The
reason being that from the entropic/energetic computation we get the
threshold condition (\ref{rot:threshold}) in the marginal unstable
case. The computation leading to this condition is exact to all
orders in the velocity, so we do not expand (\ref{Pert:Ae:rot}) that
is responsible for the unstable wavenumber cut-off in
(\ref{Pert:Dispersion:rot}).

In section \ref{sec:RegimeValidity} we will discuss the regime of
validity of the hydrodynamic description. Our results are valid for
$\{\omega R_o, kR_o \} \gg \frac{\sigma}{\rho_0 R_o}$. There is a
wide range of parameters for which this condition is satisfied. In
particular, the physically relevant most unstable mode of Figs
\ref{fig:GLstatic} and \ref{fig:GLrotate} fits very well in this
regime. Near the extreme points of the dispersion relation where
$\omega=0$ the results must be read with caution.

\subsubsection{Viscosity}

We have neglected in our analysis the effects of viscosity, which is
absent from stationary solutions but should play a role in
time-dependent processes such as the Rayleigh-Plateau instability.

These effects are well understood in the the non-relativistic
Rayleigh-Plateau instability, which has been the object of exhaustive
theoretical, numerical and experimental study. It does not seem
unreasonable to assume that the effect of viscosity in the relativistic
and non-relativistic instabilities is qualitatively similar.

Viscosity plays a subleading role in the activation of the RP
instability, although the instability can get considerably weaker if the
viscosity is very high. Generically, it increases the wavelength of the
most unstable mode, and weakens the strength of the instability
\cite{Chandrasekhar:1981}. Viscosity has been experimentally seen to
play an important role at later stages in the time evolution of the
instability, when the fluid tube pinches off, as we discuss in
sec.~\ref{sec:RPGL}

Dissipation increases considerably the technical challenge of solving
numerically the hydrodynamic perturbed equations. This is because
without it, these are ODEs with boundary conditions, while when
dissipation is present we must deal with a coupled system of second
order differential equations for the pressure and velocity
perturbations, which cannot be studied analytically and requires
advanced numerical treatment.

\section{Regime of validity \label{sec:RegimeValidity}}

We now discuss the limits of validity of the different approximations we
have made in our derivations in the previous sections.

First of all, for the fluid description of the deconfined plasma to be
accurate the thermodynamic quantities must vary slowly over the mean
free path $\ell_{\rm mfp}$ of the fluid, which is of the order of the
mass gap of the theory, or equivalently of the order of the
deconfinement temperature. That is, all length scales $\lambda$ in the
fluid must be
\eq\label{longwave}
\lambda \gg T_c^{-1}\sim \frac{\sigma}{\rho_0}\,.
\eeq
For the same
reason, thermal fluctuations must remain small. A good estimate for the valid regime
is obtained when the maximum fractional rate of change of the fluid
local temperature is small
\eq
\frac{\delta {\cal T}}{T_c}{\bigl |}_{\rm max}\sim
\partial_r \ln\gamma {\bigl |}_{\rm max}\ll 1\,.
\eeq
This occurs for
\eq
\frac{\omega_\phi^2
R_o^2}{1-\omega_\phi^2 R_o^2} \ll \frac{\rho_o R_o}{\sigma}\,.
\eeq
For
plasma balls and rings this condition is satisfied for a wide range
of lumps as long as we are away from extremality where the
temperature vanishes \cite{Lahiri:2007ae}. For plasma tubes the
angular velocity $\omega_\phi$ increases with $R_o$ so the above
condition is fulfilled as long as the tubes do not rotate too fast.
This does not affect our analysis in section \ref{sec:RProtate}
since, in order to have analytic control on the equations, we only
considered the regime of small rotation. This condition also imposes
restrictions on the validity of the Rayleigh-Plateau instability
analysis. To guarantee that the thermodynamic quantities of the
fluid vary on a length scale small when compared with the mean free
path we must demand that the Rayleigh-Plateau unstable frequencies
and wavenumbers satisfy
\eq\{\omega R_o, kR_o \} \gg
\frac{\sigma}{\rho_0 R_o}\,.
\eeq
Since the most unstable mode dominates the instability we must guarantee
that this condition is satisfied in the vicinity of the maximum of the
dispersion relation. We find that for $\frac{\sigma}{\rho_0 R_o}\lesssim
10^{-4}$ this condition is satisfied, and things get better as
$\frac{\sigma}{\rho_0 R_o}$ becomes smaller. Recall that for the
plots of Figs. \ref{fig:GLstatic} and \ref{fig:GLrotate} we took
$\frac{\sigma}{\rho_0 R_o}\sim 10^{-6}$, well within the validity range.

Second, the interface between the confined and deconfined phases,
\ie the fluid surface, has a finite thickness of the
order $1/T_c$ \cite{Aharony:2005bm}, and therefore the delta-like
surface approximation we used is valid provided that the curvature
of the surface is small with respect to $1/T_c$. So, for
plasma balls and rings the analysis is valid when the boundary
radii $\ro$, $\ri$, and $\ro-\ri$ are large compared to $T_c^{-1}$.
This is the case if the
plasma energy is large and away from the extremal
configurations \cite{Lahiri:2007ae}. For plasma tubes the requirements are,
analogously, $\ro$, $R_m$, and $\ro-R_m$ much larger than $T_c^{-1}$.

Finally, we neglected the dependence of the surface tension on the
temperature and other thermodynamic quantities of the fluid. For
consistency we must then demand that on the boundary between the
confined and deconfined phases the temperature of the plasma must
remain everywhere close to the critical temperature $T_c$. The
values of ${\cal T}$ and $T_c$ at the boundary surfaces remain close
for a large range of energies and angular momenta (see Fig.~9 of
\cite{Lahiri:2007ae}), both for balls and rings as well as for
plasma tubes, always away from the extremal limits.

\section{Rayleigh-Plateau and Gregory-Laflamme
\label{sec:RPGL}}

The dynamics of lumps of fluid that we have analyzed should admit,
via the reasoning in \cite{Bhattacharyya:2008jc}, a mapping to
the dynamics of black holes near the infrared bottom in SS-AdS, in the
limit in which the black hole size is much larger than the thickness of
the domain wall that connects it to the confining `AdS soliton' vacuum.
In this case, the correspondence between black holes and lumps of fluid
can be established and the solutions of the fluid equations can be
regarded as actual gravitational calculations in a controlled
perturbative (in boundary derivatives) expansion to the Einstein
equations in AdS, reformulated in an appropriate way.

This implies that our results for the dynamics of plasma tubes in $d$
dimensions can be regarded as describing gravitational physics of black
strings in SS-AdS$_{d+2}$ spacetime. However, such black string
solutions have not been constructed directly yet and therefore a
comparison cannot be made between results obtained using dual-fluid
techniques and more conventional gravitational analyses.

Nevertheless, it is remarkable that the dynamics of fluid tubes from our
analysis is strikingly similar to what is known for black strings in
asymptotically flat space, for which a fairly large amount of results
are available. It has been pointed out that all the different plasma
ball, pinched ball and plasma ring phases found in
\cite{Lahiri:2007ae,Bhardwaj:2008if} in $d$ dimensions have black hole,
pinched black hole, and black ring analogues in asymptotically flat
$d+2$-dimensional gravity
\cite{Emparan:2001wn,Emparan:2003sy,Emparan:2007wm}.

This parallelism extends to most aspects of the solutions we have
discussed. Ref.~\cite{Miyamoto:2008rd} already observed that the
phase diagrams for static plasma lumps (shown here in
Fig.~\ref{fig:staticphases}) resemble the corresponding phase diagrams
for black strings and black holes localized in a Kaluza-Klein circle
\cite{Kudoh:2004hs,Kol:2004ww}. Let us add that if we compare our
Fig.~\ref{fig:staticphases}.a with Fig.~3 of \cite{Harmark:2005pp}, the
two figures can be strikingly superposed. Note that in Fig.~3 of
\cite{Harmark:2005pp} the gravitational merger between the non-uniform
black string and localized black hole branches is not shown because the
available numerical code breaks in this region \cite{Kudoh:2004hs}. It
is however conjectured that the two branches do indeed converge into a
topology-changing merger with a conical structure \cite{Kol:2002xz}, and
there is numerical evidence for this \cite{Kol:2003ja}. The
approximations for the fluid description become invalid close to this
transition point (as do, too, the gravitational ones), but certainly a
topology-changing transition seems to take place when this is neglected
and the calculations are pushed until pinch-off. The similarities
between the two approaches are strong enough to lend credence to the
existence of the transition. What seems less clear is whether the
transition in the fluid involves the same conical structure as in
\cite{Kol:2002xz,Kol:2003ja}. The strong gravitational interaction
between the images of localized black holes close to the merger,
compared to the absence of interaction between the images of fluid
balls, indicates the possibility of at least differences of detail in
the two cases.

An even more uncanny similarity in the phase structure of both fluid
tubes and black strings refers to the critical dimension of spacetime
for which the transition between uniform and non-uniform phases changes
from first to second order.\footnote{In this and in the next section we
shall use $d$ and $D$, respectively, for the spacetime dimensions in
which the fluid lumps and the black holes live.} For fluid tubes
ref.~\cite{Miyamoto:2008rd} found (and we have reproduced) that this
happens at $d_*= 11$, while for black strings it happens at $D_*= 14$
for the microcanonical ensemble \cite{Sorkin:2004qq} and $D_*= 13$ for
the canonical ensemble \cite{Kudoh:2005hf}. If we assume that, as in the
AdS context, the $d$-dimensional fluids are to be related to
$D=d+2$-dimensional black holes, then the apparent closeness in the
values of the critical dimensions is truly startling.

The qualitative similarities between the dispersion relations in the
Rayleigh-Plateau and in the Gregory-Laflamme instability have also been
pointed out before for the case of static tubes and strings
\cite{Cardoso:2006ks}. The parallels extend to further dynamical
details, such as the fact that for static solutions, non-axisymmetric
modes ($m\neq 0$) are stable in the plasma, while the GL instability
also only affects $s$-modes \cite{Hovdebo:2006jy,Kudoh:2006bp}.
Furthermore, a boost along the static plasma tube has the same effect as
boosting the black string: it is only a kinematic effect increasing the
threshold wavenumber for $s$-modes and leaving all other modes stable
\cite{Hovdebo:2006jy}.

Rotating black strings are affected by the
same instability. The GL instability in rotating black strings was
analysed in \cite{Kleihaus:2007dg}. They concluded that the GL
instability persists for rotating black strings all the way to
extremality and the threshold wavenumber increases as the rotation grows
(see Fig.~1 of \cite{Kleihaus:2007dg}). We could only treat small
rotation analytically, but our dual plasma results, summarized in
Fig.~\ref{fig:GLrotate}, agree with this: the instability persists as
rotation grows and it actually gets stronger, with the critical
wavenumber increasing with the rotation. In fact this can be understood
intuitively. The centrifugal force scales with the radial distance, so
it is larger at the crest of the non-uniformity triggered by the RP
instability than at the trough. As a result the distance between the
crest and trough is stretched: rotation enhances the instability. An
interesting qualitatively new feature that rotation adds is that
non-axisymmetric modes ($m\neq 0$) can become unstable to the
Rayleigh-Plateau instability for sufficiently large rotation.

Among the phases for rotating tubes that we have uncovered, a
particularly interesting one is the pinched non-uniform tube (pNUT),
since it has no analogue for static tubes. Its shape certainly
suggests that we should think of it as the result of a bulge in the
non-uniform tube beginning to behave like a rotating ball that
develops a pinch. Since pinched black holes are also expected to
exist in six-dimensional vacuum gravity (actually in any dimension
larger than five) \cite{Emparan:2003sy,Emparan:2007wm}, it is then
natural to conjecture that rotating black strings in six dimensions
also have phases of pinched non-uniform black strings, in addition
to the already known uniform black strings, uniform black tubes (\ie
the black ring of \cite{Emparan:2001wn} times a flat direction), and
the expected more conventional non-uniform black string.

Such remarkable similitude between the dynamics of fluid tubes and
of vacuum black strings makes it tempting to try to push it further
in order to address an outstanding open problem: the time evolution
of the GL instability and its final fate. The final stage of this
evolution is not known in gravity, since the available numerical
code, which has been developed only for $D=5$, breaks down for late times
\cite{Choptuik:2003qd}. On the other hand, the time evolution of the
(non-relativistic) RP instability has been the object of several
experimental and numerical studies, see \eg \cite{Stone} which
considers viscous fluids in four spacetime dimensions
(unfortunately, higher-dimensional fluids remain unavailable
experimentally). We would expect that for static initial data, the
classical and relativistic evolution should not differ
substantially. It is then justified to compare the initial time
evolution of the GL instability, Fig.~4 of \cite{Choptuik:2003qd},
with the early stage dynamical evolution of the RP instability, \eg
Fig.~1 of \cite{Stone}. There appears to be a nice match between the
two. In the fluid, where the evolution has been followed until its
endpoint, the uniform tube pinches-off and breaks. Starting from a
single sinusoidal perturbation in a cylindrical liquid bridge (fluid
tube), the higher harmonics generated by non-linear effects are
responsible for the development of a long neck that breaks the tube
in a self-similar process. One ends up with an array of satellite
drops with different sizes \cite{Stone}\footnote{Actually, two main
regimes have been observed depending on the viscosity of the fluid.
For low viscosity, there is repeated stretching and breakup, ending
with several satellite droplets. For high viscosity, a single
breakup that takes a long time can occur and in the end a single
satellite drop is formed.}.

We may expect such behavior to be representative of the time evolution
of the instability of fluids in subcritical dimensions $d<d_*=11$. A
numerical study of the non-relativistic evolution for inviscid fluids in
several dimensions has been made in \cite{Miyamoto:2008uf}. The endpoint
of the RP instability does appear to differ for subcritical and
supercritical dimensions. For $d<d_*$ the endpoint is a drop or an array
of drops. Again, it is tempting to argue in favor of a similar evolution
in the gravitational system, in which the black string pinches off
completely and forms an array of black holes, possibly of different
sizes\footnote{These have already been constructed in
\cite{Dias:2007hg}. It is entropically favorable for these arrays of
black holes to merge into a single one.}. The same caveats apply here as
in the discussion of the ball-tube/black hole-black string transition:
the fluid description breaks down near pinch-off; and blobs of fluid do
not self-interact whereas gravitating horizon blobs do. However, it is
already significant that pinch off does occur for viscous fluids, and
that no study reveals any sign of the instability slowing down as
pinch-off is approached in $d<d_*$.

The situation may be cleaner above the critical dimension.
Ref.~\cite{Miyamoto:2008uf} finds that in this case the endpoint of the
RP instability is a non-uniform tube with constant mean curvature. In
this case there is no concern about the validity of the fluid
description, since the tube can remain thick enough during its evolution.
The role of viscosity is not too clear, but it might be expected to
smooth the evolution. This suggests that the endpoint of the GL
instability on a black string above the critical dimension could be a
non-uniform black string, like ref.~\cite{Horowitz:2001cz} suggested.

The non-linear GL evolution when rotation is present has not been
studied yet but again, the (non-relativistic) RP evolution of fluids is
well studied (see \eg \cite{KubWei}): rotation introduces twisting
effects in the pinch off of the tube and as rotation increases a
dramatic centrifugal ejection of drops is observed (see the photographs
of \cite{KubWei}).

Of course it is possible that black strings and black holes in vacuum
gravity behave in a closely similar way to fluids only as long as one considers
configurations that are stationary, slowly evolving, and/or not too
non-uniform. The dynamics of vacuum gravity may depart significantly
from fluid dynamics away from these regimes and furthermore, in
contrast to the situation for gravity in AdS, it is not obvious whether
it can be accurately pictured as fluid dynamics in some controlled
approximation. The next section addresses this issue in more
detail.


\section{The black hole/fluid analogy revisited}
\label{sec:revisit}

The similarities discussed in the previous section naturally prompt the
question of to what extent we can trust fluid dynamics as a guide for
unknown dynamics of {\it vacuum} black holes in asymptotically flat
space. In other words, since the analogy between
fluids and black holes was observed in the first place for
asymptotically flat black holes, is there any precise meaning to it?
Clearly, we cannot just take the results for AdS black holes in the
limit in which the cosmological constant vanishes. The fluid description
for AdS black holes requires that their size be much larger than the AdS
curvature radius, so black holes smaller than this size fall outside the
scope of the dual hydrodynamics.

We have discussed a number of
qualitative or semi-quantitative features that fluid lumps share with
black holes in vacuum gravity. However, there
is at least one crucial difference that would appear to preclude a
precise correspondence. Classical General Relativity in vacuum is a
scale-invariant theory, which implies that all properties of a black
hole scale uniformly with mass. That is, all Schwarzschild black
holes are essentially the same; all Kerr black holes with the same
value of $J/GM^2$ also have the same properties; and all black
strings in Kaluza-Klein space are the same when the compact radius
is scaled to keep $L/(GM)^{1/(D-2)}$ constant, \eg the
Gregory-Laflamme threshold mode wavelength for a static black string
must scale like the inverse of the horizon transverse radius. This
scale invariance is not present for the fluid: the surface tension
sets a scale that distinguishes fluid lumps of different size: a
fluid ball of a given radius is not simply a scaled-up version of a
ball of half that radius ---for instance, the former can more easily
break up than the latter. For a large plasma ball, the relative
entropy cost of breaking it in two pieces becomes arbitrarily small
as the radius $R$ of the ball gets arbitrarily large. For instance,
for a ball of fluid with initial entropy $S_i$, temperature $T_i$ and radius $R_i$, that breaks
into two equal balls, each with entropy $S_f/2$, temperature $T_f$ and
radius $R_f$,
keeping the total energy unchanged, the ratio between final and initial
entropies is
\beq \label{breakball1}
\left.\frac{S_f}{S_i}\right|_\mathrm{fluid}\simeq
\frac{T_i}{T_f}\left(1+\frac{\sigma}{\rho}\left(\frac{1}{R_i}-
\frac{1}{R_f}\right)\right)\,,
\eeq
where we have assumed that the ball radii are much larger than the
length scale $\sigma/\rho$ set by the energy density
$\rho$ and surface tension $\sigma$ of the fluid\footnote{To leading
order we can neglect the difference in energy density in the initial and
final fluid balls.}.
To leading order, $R_f^{-1}=2^\frac{1}{d-1}R_i^{-1}>R_i^{-1}$, and moreover,
if the balls have negative
specific heat the smaller final balls will be hotter than the initial
one, $T_f>T_i$, with $T_f/T_i-1$ a positive and small quantity of
order $\sigma/(\rho R_i)$. Then $S_f<S_i$, so the
process is
suppressed, but the relative entropy cost of breaking
\beq \label{breakball}
\left.\frac{\Delta S}{S}\right|_\mathrm{fluid}=\left.\frac{S_f-S_i}{S_i}\right|_\mathrm{fluid}\simeq
-\delta_d\frac{\sigma}{\rho R_i}
\eeq
is small, with large fluid balls breaking more easily than smaller ones.
Here $\delta_d$ is a positive, dimension-dependent number which, for a
plasma with the equation of state considered in
sec.~\ref{sec:StationaryPlasma}, is
\beq\label{breakball3}
\delta_d=(d-1)\left(2^{\frac{1}{d-1}}-1\right)\,.
\eeq
%
%

%
For a
Schwarzschild black hole, in contrast, the relative entropy cost for
splitting into two equal black holes keeping the total mass fixed,
remains constant independently
of the black hole size,
\beq \label{breakbh}
\left.\frac{\Delta S}{S}\right|_\mathrm{bh}=\left.\frac{S_f-S_i}{S_i}\right|_\mathrm{bh}=
2^{-\frac{1}{D-3}}-1\,.
\eeq
It is amusing that, if we make the identification $D=d+2$ that is
suggested by the duality between SS-AdS$_{d+2}$ black holes and
$d$-dimensional fluid lumps, the powers of $2$ in these formulas
coincide. But on the other hand, the identification between
\eqref{breakball} and \eqref{breakbh} appears to require that
$\sigma\sim \rho R$. This is not only a rather bizarre behavior for
the surface tension, which grows linearly with the size of the fluid
ball. It also implies that the black hole is equated with a fluid ball
that is always small\footnote{This is indeed the case for small black
holes in AdS, which resemble asymptotically flat ones.}, in the sense
that surface effects are always strong so the regime \eqref{longwave} is
never attained. The apparent reason for this is, again, that vacuum
gravity does not have a scale parameter to characterize a black hole as
parametrically large.

So the crux of the problem is that we are trying to relate objects
in parameter spaces of different dimensionality! For the black holes, as
argued, we are always free to fix a scale so as to set the mass, or the
length $(GM)^\frac{1}{D-3}$, equal to one, and then completely characterize the black
hole by its angular momenta. For a fluid lump, the theory comes with a
scale already, namely the length scale $\sigma/\rho$, so in order to specify a
fluid ball we must provide its size relative to that scale (or its
energy), and its angular momenta --- \ie one more parameter than for a
black hole.

One may point out that a natural scale that appears in black hole
physics is the Planck length, which does allow to make a distinction
between small and large black holes. However, this is of no help to make
a better comparison between vacuum black holes and fluid lumps, since in
the limit in which the `cutoff' scale becomes negligible, namely,
when the fluid lumps or the black holes become very large relative to
that scale, their properties are clearly different, as illustrated
above. It would be desirable to have a way of deciding whether a black
hole is large or small within {\it classical} vacuum General Relativity,
but this seems hard in the absence of a classical fundamental scale.

Does this, then, imply, that there can be no limit in which the equations of
the dynamics and the phase space of {\it vacuum} black holes can be mapped into the
equations and phase space of a fluid, with effective
parameters that satisfy
\beq
\frac{\sigma}{\rho R}\ll 1\,?
\eeq
 In the following we make some observations
that suggest otherwise.

Let us first note that an obvious difference between fluids and black
holes is that two disconnected lumps of fluid do not attract each other
(at least within the framework in which we are considering fluids). This
must obviously imply that some phenomena, like the deep non-linear
evolution of the GL instability, or the mergers involving vacuum black
holes or black rings, may behave differently than those for their fluid
counterparts: in the former case the gravitational attraction between
different lumps on the horizon presumably plays a role similar to the
attraction between two black holes and thus affects their
evolution\footnote{That these effects must, of necessity, be strongly
suppressed for the class of AdS black holes that are dual to plasma
lumps, is a remarkable consequence of the duality.}. Thus, any limit in
which vacuum black holes behave accurately as fluids must be a limit in
which this gravitational attraction is suppressed.

Such a limit seems to arise if we consider gravity in a spacetime with a
large number of non-compact dimensions. As the number of dimensions
increases, the gravitational potential becomes steeper and more
localized near the source, and flatter and weaker at larger distances.
Furthermore, since the black hole entropy
\beq
S\propto M^{\frac{D-2}{D-3}}
\eeq
is directly proportional to the mass in the limit $D\to\infty$, there is
no entropy cost in splitting a black hole: \eqref{breakbh} yields
$\Delta S/S\to 0$ in this limit. As we observed above, this is just
like for a large drop of fluid. Indeed, both for a fluid with
vanishingly small surface tension and for a black hole in $D\to\infty$
the (Hagedorn-like) relation $E=TS$ is satisfied\footnote{For the fluid
lump $E$ here is the difference between its energy and the energy
of the (unconfined) vacuum it displaces.}.

This enhanced instability of the black hole horizon to break up has a
rather precise manifestation in the equivalence between the
Gregory-Laflamme and Rayleigh-Plateau instabilities. Given the
observations above, we can expect a black string in $D\to\infty$ to be
able to split into fragments of any length, \ie the wavelength of the
threshold mode is expected to approach zero. Indeed, the GL instability
in arbitrary dimension has been studied in \cite{Kol:2004pn}, and even if the
equation for the threshold mode is fairly complicated, it simplifies
greatly if we expand in $1/D$ to leading order,
\beq\label{thrGL}
\chi''(r)+ \frac{D}{r}\chi'(r)-k^2\chi(r)=0\,.
\eeq
This is essentially the same as the equation for the RP threshold
($\omega=0$) mode \eqref{Pert:Bes:st}, and it becomes precisely the same
if we identify $D\simeq n\simeq d$ for large $D$ and $d$. The wavelength
of the GL threshold mode in this limit vanishes as $r_s/\sqrt{D}$, in
accord with the expectation above.

Expanding on these observations, we can try to identify what plays the
role of the dimensionless quantity $\sigma/(\rho R)$ in the gravitational
side. The answer is obvious: since the only small dimensionless
parameter we have at our disposal in the gravity side is $1/D$, both
must be related. A more precise mapping can be obtained by comparing
\eqref{breakball} and \eqref{breakbh}. For large $D$, the latter becomes
\eq
\left. \frac{\Delta S}{S}\right|_\mathrm{bh}\simeq -\frac{\log{2}}{D}\,,
\eeq
which leads to the identification
\eq\label{sigmaD}
\frac{\sigma}{\rho R}\sim \frac{1}{D}\,.
\eeq
We are not necessarily proposing that the limit $d\to\infty$ must be
taken in the fluid side too, but if we do, and use the value
\eqref{breakball3}, $\delta_d\to \log 2$, then the identification
becomes precise,
$\frac{\sigma}{\rho R}= D^{-1}$,
at least to leading order in $1/D$.

Eq.~\eqref{sigmaD} illustrates a main feature of our proposal. As we
have discussed, the main problem for a fluid/black hole correspondence
is the apparent absence in classical vacuum GR of a way to decide
whether a black hole is intrinsically large or small. Our suggestion
amounts to saying that {\it a black hole is large or small depending on
the number of spacetime dimensions it lives in.} Black holes in low
dimensions are to be regarded as tiny droplets, with surface
energy comparable to bulk energy, whereas black holes in very high
dimensions behave rather like large lumps of fluid. In other words, we
are using $D$ as an additional parameter characterizing a black hole.
We are making
crucial use here of the observation that in classical General Relativity
in vacuum the only dimensionless tunable parameter at our disposal is
$D$, a point that has also been urged by B.~Kol and collaborators
\cite{Asnin:2007rw}.

It might be that these similarities between fluids and black holes at
large $D$ only reflect purely geometrical aspects of the relation
between volumes and areas at very high dimensions (note in particular
that $D$, but not $\sigma/(\rho R)$, appears in \eqref{thrGL}). But at
any rate we feel that these observations suggest that perhaps by
considering a $1/D$ expansion, and possibly accounting for the effects
of gravitational interaction in this expansion, a more accurate map
betwen vacuum black holes and fluids may be achieved that allows to
understand not only the similarities that have been observed, but also
the differences.


\section*{Acknowledgments}

We thank Vitor Cardoso,  Veronika Hubeny, Umpei Miyamoto, Renaud
Parentani and Toby Wiseman for interesting discussions. MMC, OJCD
and RE were supported in part by DURSI 2005 SGR 00082, MEC FPA
2004-04582-C02 and FPA-2007-66665-C02, and the European Community
FP6 program MRTN-CT-2004-005104. OJCD and RE thank the Niels Bohr
Institute for hospitality and the organizers of the workshop
``Mathematical Aspects of General Relativity", Copenhagen, April
2008; OJCD, RE and DK thank CERN for hospitality during the
programme ``Black Holes: A Landscape of Theoretical Physics
Problems", August-October 2008; and MMC, OJCD and RE are grateful to
the organizers of the workshop ``Quantum Black Holes, Braneworlds
and Holography", Valencia, May 2008, where part of this work was
done. MMC aknowledges financial support provided by the FWO\,-\,Vlaanderen, project G.0235.05 and in part by the Federal Office for ScientiÞc, Technical and Cultural Affairs through the ÔInteruniversity Attraction Poles Programme Ð Belgian Science PolicyÕ P6/11-P. 
OJCD acknowledges financial support provided by the European
Community through the Intra-European Marie Curie contract
MEIF-CT-2006-038924. This work was partially funded by FCT through
project PTDC/FIS/64175/2006. DK was supported in part by INFN,
MIUR-PRIN contract 20075ATT78, and by the European Community FP6
program MRTN-CT-2004-005104.



\begin{thebibliography}{99}

\bibitem{Cardoso:2006ks}
  V.~Cardoso and O.~J.~C.~Dias,
  ``Gregory-Laflamme and Rayleigh-Plateau instabilities,''
  Phys.\ Rev.\ Lett.\  {\bf 96} (2006) 181601
  [arXiv:hep-th/0602017].

\bibitem{Cardoso:2007ka}
  V.~Cardoso, O.~J.~C.~Dias and L.~Gualtieri,
  ``The return of the membrane paradigm? Black holes and strings in the   water
  tap,''
  Int.\ J.\ Mod.\ Phys.\  D {\bf 17} (2008) 505
  [arXiv:0705.2777 [hep-th]].

\bibitem{Cardoso:2006sj}
  V.~Cardoso and L.~Gualtieri,
  ``Equilibrium configurations of fluids and their stability in higher
  dimensions,''
  Class.\ Quant.\ Grav.\  {\bf 23} (2006) 7151
  [arXiv:hep-th/0610004].

\bibitem{Gregory:1993vy}
  R.~Gregory and R.~Laflamme,
  ``Black strings and p-branes are unstable,''
  Phys.\ Rev.\ Lett.\  {\bf 70} (1993) 2837
  [arXiv:hep-th/9301052].

\bibitem{Gregory:1994bj}
  R.~Gregory and R.~Laflamme,
  ``The Instability of charged black strings and p-branes,''
  Nucl.\ Phys.\  B {\bf 428} (1994) 399
  [arXiv:hep-th/9404071].

\bibitem{Bhattacharyya:2007vs}
   S.~Bhattacharyya, S.~Lahiri, R.~Loganayagam and S.~Minwalla,
   ``Large rotating AdS black holes from fluid mechanics,''
   arXiv:0708.1770 [hep-th].

\bibitem{Bhattacharyya:2008jc}
  S.~Bhattacharyya, V.~E.~Hubeny, S.~Minwalla and M.~Rangamani,
  ``Nonlinear Fluid Dynamics from Gravity,''
  JHEP {\bf 0802} (2008) 045
  [arXiv:0712.2456 [hep-th]].

\bibitem{Copsey:2006br}
  K.~Copsey and G.~T.~Horowitz,
  ``Gravity dual of gauge theory on S**2 x S**1 x R,''
  JHEP {\bf 0606}, 021 (2006)
  [arXiv:hep-th/0602003].

\bibitem{Mann:2006yi}
  R.~B.~Mann, E.~Radu and C.~Stelea,
  ``Black string solutions with negative cosmological constant,''
  JHEP {\bf 0609} (2006) 073
  [arXiv:hep-th/0604205].

\bibitem{Bernamonti:2007bu}
  A.~Bernamonti, M.~M.~Caldarelli, D.~Klemm, R.~Olea, C.~Sieg and E.~Zorzan,
  ``Black strings in AdS$_5$,''
  JHEP {\bf 0801} (2008) 061
  [arXiv:0708.2402 [hep-th]].

\bibitem{Brihaye:2007ju}
  Y.~Brihaye, T.~Delsate and E.~Radu,
  ``On the stability of AdS black strings,''
  Phys.\ Lett.\  B {\bf 662} (2008) 264
  [arXiv:0710.4034 [hep-th]].

\bibitem{Aharony:2005bm}
  O.~Aharony, S.~Minwalla and T.~Wiseman,
  ``Plasma-balls in large N gauge theories and localized black holes,''
  Class.\ Quant.\ Grav.\  {\bf 23}, 2171 (2006)
  [arXiv:hep-th/0507219].

\bibitem{Lahiri:2007ae}
  S.~Lahiri and S.~Minwalla,
  ``Plasmarings as dual black rings,''
  arXiv:0705.3404 [hep-th].

\bibitem{Bhardwaj:2008if}
  S.~Bhardwaj and J.~Bhattacharya,
  ``Thermodynamics of Plasmaballs and Plasmarings in 3+1 Dimensions,''
  arXiv:0806.1897 [hep-th].

\bibitem{Emparan:2008eg}
  R.~Emparan and H.~S.~Reall,
  ``Black Holes in Higher Dimensions,''
  Living Rev.\ Rel.\  {\bf 11} (2008) 6
  [arXiv:0801.3471 [hep-th]].

\bibitem{Kol:2004ww}
  B.~Kol,
  ``The phase transition between caged black holes and black strings: A
  review,''
  Phys.\ Rept.\  {\bf 422} (2006) 119
  [arXiv:hep-th/0411240].

\bibitem{Obers:2008pj}
  N.~A.~Obers,
  ``Black Holes in Higher-Dimensional Gravity,''
  arXiv:0802.0519 [hep-th].

\bibitem{Gubser:2001ac}
S.~S. Gubser, ``On non-uniform black branes,'' Class.\ Quant.\
Grav.\ {\bf 19} (2002) 4825, [arXiv:hep-th/0110193].

\bibitem{Wiseman:2002zc}
  T.~Wiseman,
  ``Static axisymmetric vacuum solutions and non-uniform black strings,''
  Class.\ Quant.\ Grav.\  {\bf 20}, 1137 (2003)
  [arXiv:hep-th/0209051].


\bibitem{Miyamoto:2008rd}
  U.~Miyamoto and K.~i.~Maeda,
  ``Liquid bridges and black strings in higher dimensions,''
  Phys.\ Lett.\  B {\bf 664}, 103 (2008)
  [arXiv:0803.3037 [hep-th]].

\bibitem{Miyamoto:2008uf}
  U.~Miyamoto,
  ``Curvature driven diffusion, Rayleigh-Plateau, and Gregory-Laflamme,''
  Phys.\ Rev.\  D {\bf 78} (2008) 026001
  [arXiv:0804.1723 [hep-th]].

\bibitem{Dias:2007hg}
  O.~J.~C.~Dias, T.~Harmark, R.~C.~Myers and N.~A.~Obers,
  ``Multi-black hole configurations on the cylinder,''
  Phys.\ Rev.\  D {\bf 76}, 104025 (2007)
  [arXiv:0706.3645 [hep-th]].

\bibitem{mandm}
K.~i.~Maeda and U.~Miyamoto,
``Black hole-black string phase transitions from hydrodynamics,''
arXiv:0811.2305 [hep-th].


\bibitem{Bhattacharyya:2008ji}
  S.~Bhattacharyya, R.~Loganayagam, S.~Minwalla, S.~Nampuri, S.~P.~Trivedi and S.~R.~Wadia,
  ``Forced Fluid Dynamics from Gravity,''
  arXiv:0806.0006 [hep-th].

\bibitem{VanRaamsdonk:2008fp}
  M.~Van Raamsdonk,
  ``Black Hole Dynamics From Atmospheric Science,''
  JHEP {\bf 0805} (2008) 106
  [arXiv:0802.3224 [hep-th]].

\bibitem{young}
T.~Young, ``An essay on the cohesion of fluids,'' Philos. Trans. R.
Soc. London {\bf 95}, 65 (1805).

P.~S.~Laplace, ``Trait\'e de M\'ecanique C\'eleste; Suppl\'ement au
Dixi\`eme Livre, Sur l'Action Capillaire,'' (Courcier, Paris, 1806);
``Suppl\'ement \`a la Th\'eorie de l'Action Capillaire,'' (Courcier,
Paris, 1806).

\bibitem{Harmark:2005pp}
  T.~Harmark and N.~A.~Obers,
  ``Phases of Kaluza-Klein black holes: A brief review,''
  arXiv:hep-th/0503020.


\bibitem{Horowitz:2002dc}
  G.~T.~Horowitz,
  ``Playing with black strings,''
  arXiv:hep-th/0205069.

\bibitem{Hovdebo:2006jy}
  J.~L.~Hovdebo and R.~C.~Myers,
  ``Black rings, boosted strings and Gregory-Laflamme,''
  Phys.\ Rev.\  D {\bf 73} (2006) 084013
  [arXiv:hep-th/0601079].

\bibitem{Chandrasekhar:1981}
S.~Chandrasekhar, ``Hydrodynamic and Hydromagnetic Stability,''
Dover, New York, 1981.

\bibitem{Emparan:2001wn}
  R.~Emparan and H.~S.~Reall,
  ``A rotating black ring in five dimensions,''
  Phys.\ Rev.\ Lett.\  {\bf 88} (2002) 101101
  [arXiv:hep-th/0110260].

R.~Emparan and H.~S.~Reall,
  ``Black rings,''
  Class.\ Quant.\ Grav.\  {\bf 23} (2006) R169
  [arXiv:hep-th/0608012].

\bibitem{Emparan:2003sy}
  R.~Emparan and R.~C.~Myers,
  ``Instability of ultra-spinning black holes,''
  JHEP {\bf 0309} (2003) 025
  [arXiv:hep-th/0308056].

\bibitem{Emparan:2007wm}
  R.~Emparan, T.~Harmark, V.~Niarchos, N.~A.~Obers and M.~J.~Rodriguez,
  ``The Phase Structure of Higher-Dimensional Black Rings and Black Holes,''
  JHEP {\bf 0710} (2007) 110
  [arXiv:0708.2181 [hep-th]].

\bibitem{Kudoh:2004hs}
H.~Kudoh and T.~Wiseman, ``Connecting black holes and black
strings,''
  Phys. Rev. Lett. {\bf 94} (2005) 161102,
arXiv:hep-th/0409111.

\bibitem{Kol:2002xz}
  B.~Kol,
  ``Topology change in general relativity and the black-hole black-string
  transition,''
  JHEP {\bf 0510}, 049 (2005)
  [arXiv:hep-th/0206220].

\bibitem{Kol:2003ja}
  B.~Kol and T.~Wiseman,
  ``Evidence that highly non-uniform black strings have a conical waist,''
  Class.\ Quant.\ Grav.\  {\bf 20}, 3493 (2003)
  [arXiv:hep-th/0304070].

\bibitem{Sorkin:2004qq}
E.~Sorkin, ``A critical dimension in the black-string phase
transition,'' {\em
  Phys. Rev. Lett.} {\bf 93} (2004) 031601,
arXiv:hep-th/0402216.

\bibitem{Kudoh:2005hf}
  H.~Kudoh and U.~Miyamoto,
  ``On non-uniform smeared black branes,''
  Class.\ Quant.\ Grav.\  {\bf 22} (2005) 3853, arXiv:hep-th/0506019.

\bibitem{Kudoh:2006bp}
  H.~Kudoh,
  ``Origin of black string instability,''
  Phys.\ Rev.\  D {\bf 73} (2006) 104034
  [arXiv:hep-th/0602001].

\bibitem{Kleihaus:2007dg}
  B.~Kleihaus, J.~Kunz and E.~Radu,
  ``Rotating nonuniform black string solutions,''
  JHEP {\bf 0705} (2007) 058
  [arXiv:hep-th/0702053].

\bibitem{Choptuik:2003qd}
  M.~W.~Choptuik, L.~Lehner, I.~Olabarrieta, R.~Petryk, F.~Pretorius and H.~Villegas,
  ``Towards the final fate of an unstable black string,''
  Phys.\ Rev.\  D {\bf 68} (2003) 044001
  [arXiv:gr-qc/0304085].

\bibitem{Stone} M.~Tjahjadi, H.~A.~Stone and J.~M.~Ottino, ``Satellite and
subsatellite formation in cappilary breakup,'' J. Fluid Mech. {\bf
243} (1992) 297.

\bibitem{Horowitz:2001cz}
  G.~T.~Horowitz and K.~Maeda,
  ``Fate of the black string instability,''
  Phys.\ Rev.\ Lett.\  {\bf 87} (2001) 131301
  [arXiv:hep-th/0105111].


\bibitem{KubWei} J. P. Kubitschek and P. D. Weidman,
``Helical instability of a rotating viscous liquid jet,'' Phys.
Fluids {\bf 19} (2007) 114108.

\bibitem{Kol:2004pn}
  B.~Kol and E.~Sorkin,
  ``On black-brane instability in an arbitrary dimension,''
  Class.\ Quant.\ Grav.\  {\bf 21} (2004) 4793
  [arXiv:gr-qc/0407058].

  T.~Harmark, V.~Niarchos and N.~A.~Obers,
  ``Instabilities of black strings and branes,''
  Class.\ Quant.\ Grav.\  {\bf 24} (2007) R1
  [arXiv:hep-th/0701022].



\bibitem{Asnin:2007rw}
  V.~Asnin, D.~Gorbonos, S.~Hadar, B.~Kol, M.~Levi and U.~Miyamoto,
  ``High and Low Dimensions in The Black Hole Negative Mode,''
  Class.\ Quant.\ Grav.\  {\bf 24} (2007) 5527
  [arXiv:0706.1555 [hep-th]].





\end{thebibliography}
\end{document}